\documentclass[%
 reprint,
preprintnumbers,
nofootinbib,
 amsmath,amssymb,
 aps,
]{revtex4-2}

\usepackage[colorlinks=true, pdfstartview=FitV, linkcolor=magenta,citecolor=blue, urlcolor=magenta,
bookmarks=true, bookmarksnumbered=true, breaklinks]{hyperref}
\usepackage[dvipdfmx]{graphicx}
\usepackage{here}
\usepackage{url}
\usepackage{graphicx}
\usepackage{dcolumn}
\usepackage{bm}
\usepackage{comment}

\newcommand{\TTRR}{\mathbb{T}^2\times\mathbb{R}^2}
\newcommand{\SRRR}{\mathbb{S}^1\times\mathbb{R}^3}

\begin{document}

\preprint{YITP-24-47, J-PARC-TH-0304}

\title{
Novel first-order phase transition and critical points in SU(3) Yang-Mills theory \\ with spatial compactification
}

\author{Daisuke~Fujii}
\email[]{daisuke@rcnp.osaka-u.ac.jp}
\email[]{fujii.daisuke@jaea.go.jp}
\affiliation{%
Advanced Science Research Center, Japan Atomic Energy Agency, Tokai, Ibaraki 319-1195 Japan
}%
 \affiliation{Research Center for Nuclear Physics,
Osaka University, Ibaraki 567-0048, Japan}

\author{Akihiro~Iwanaka}%
 \email[]{iwanaka@rcnp.osaka-u.ac.jp}
 \affiliation{Research Center for Nuclear Physics,
Osaka University, Ibaraki 567-0048, Japan }

\author{Masakiyo~Kitazawa}
\email[]{kitazawa@yukawa.kyoto-u.ac.jp}
\affiliation{Yukawa Institute for Theoretical Physics, Kyoto University, Kyoto, 606-8502 Japan}
\affiliation{J-PARC Branch, KEK Theory Center, KEK, Tokai, 319-1106 Japan}

\author{Daiki~Suenaga}
\email[]{suenaga.daiki.j1@f.mail.nagoya-u.ac.jp}
\affiliation{Kobayashi-Maskawa Institute for the Origin of Particles and the Universe, Nagoya University, Nagoya, 464-8602, Japan}
\affiliation{Research Center for Nuclear Physics,
Osaka University, Ibaraki 567-0048, Japan}

\date{\today}

\begin{abstract}
We investigate the thermodynamics and phase structure of $SU(3)$ Yang-Mills theory on $\mathbb{T}^2\times\mathbb{R}^2$ in Euclidean spacetime in an effective-model approach. The model incorporates two Polyakov loops along two compactified directions as dynamical variables, and is constructed to reproduce thermodynamics on $\mathbb{T}^2\times\mathbb{R}^2$ measured on the lattice. The model analysis indicates the existence of a novel first-order phase transition on $\mathbb{T}^2\times\mathbb{R}^2$ in the deconfined phase, which terminates at critical points that should belong to the two-dimensional $Z_2$ universality class. We argue that the interplay of the Polyakov loops induced by their cross term in the Polyakov-loop potential is responsible for the manifestation of the first-order transition.
\end{abstract}

\maketitle


\section{\label{sec:level1}introduction}

Thermodynamic quantities, such as pressure and energy density, are fundamental observables characterizing properties of a medium in equilibrium.
In Quantum Chromodynamics (QCD) at zero baryon density and pure Yang-Mills (YM) theories, thermodynamics measured in numerical simulations on the lattice~\cite{Boyd:1996bx,Umeda:2008bd,Asakawa:2013laa,Borsanyi:2012ve,Borsanyi:2013bia,HotQCD:2014kol,Taniguchi:2016ofw,Kitazawa:2016dsl,Giusti:2016iqr,Caselle:2018kap,Iritani:2018idk} have played central roles in revealing nontrivial properties of these theories at nonzero temperature, such as the rapid crossover around the pseudo-critical temperature in QCD at physical quark masses.
Detailed understanding of the QCD thermodynamics is also indispensable to investigate the hot medium created by the relativistic heavy-ion collisions~\cite{Yagi:2005yb}. 

While spatial isotropy is typically assumed, thermodynamics can accommodate systems that are spatially anisotropic. Such systems are realized, for example, by imposing boundary conditions (BCs). A well-known example is the Casimir effect~\cite{Casimir:1948dh}, where the BC imposed by conductors gives rise to an anisotropic pressure~\cite{Brown:1969na}. 
Studying anisotropic systems may also be important in understanding the evolution of fireballs created by relativistic heavy-ion collisions, which are not only finite-volume systems but also have highly distorted shapes. 

Among thermal systems having spatial anisotropy, systems with a periodic boundary condition (PBC) in one spatial direction, i.e. a spatial compactification,  are simple but nontrivial examples. These systems possess translational symmetry that makes theoretical and numerical treatments tractable, while the pressure along the compactified and uncompactified directions can generally be different, which adds a new thermodynamic quantity into the system. Since the compactified length is a new parameter to specify the system, the variables controlling the system are also increased. These degrees of freedom can be exploited as new probes to unveil the system's properties.

Properties of various theories in this setup have been investigated from various motivations~\cite{Hanada:2009hq,Hanada:2010kt,Unsal:2010qh,Mandal:2011hb,Mandal:2013id,Ishikawa:2018yey,Ishikawa:2019dcn,Kitazawa:2019otp,Inagaki:2021yhi,Chernodub:2022izt,Tanizaki:2022plm,Tanizaki:2022ngt,Hayashi:2023wwi,Hayashi:2024qkm}. In Ref.~\cite{Kitazawa:2019otp}, anisotropic thermodynamics in $SU(3)$ YM theory has been investigated in lattice numerical simulations for temperatures near but above the critical temperature of the deconfinement transition $T_{\rm d}$. It was found that the anisotropy in the pressure arising from the BC in this theory is significantly suppressed compared to that in the free boson theory. Revealing its physical origin is an interesting subject 
that will enrich our understanding of the non-perturbative nature of this theory from a new perspective.

In the Matsubara formalism, thermal bosonic systems at temperature $T$ are represented by the Euclidean spacetime with the PBC along the temporal direction of length $1/T$. In $SU(N)$ YM theory, it is known that the center $Z_N$ symmetry associated with the BC is responsible for the confinement phase transition at nonzero $T$.
The relation between the dynamics of the spontaneous breaking of this 
symmetry and thermodynamics has also been discussed in the literature~\cite{Meisinger:2001cq,Pisarski:2000eq,Dumitru:2010mj,Dumitru:2012fw,Fukushima:2017csk}. In Ref.~\cite{Meisinger:2001cq}, effective models including the Polyakov loop, an order parameter of the $Z_N$ symmetry breaking, as a dynamical variable have been employed to discuss their relation. It has been found that the simple effective models are capable of qualitatively reproducing both the characteristic behaviors of thermodynamic quantities and the gradual growth of the Polyakov loop near $T_{\rm d}$ measured on the lattice simultaneously. This idea has been refined later in Refs.~\cite{Dumitru:2010mj,Dumitru:2012fw} by improving the model with an emphasis on the behavior of the interaction measure observed on the lattice~\cite{Pisarski:2000eq}.

When a spatial compactification is imposed into a thermal bosonic system, it is described by the Euclidean spacetime on $\TTRR$ with two PBCs~\footnote{The $\mathbb{T}^2$ that considered in this study is not general reparametrization invariant. We impose the BC orthogonal to the $x,\tau$ direction to consider the same situation as in the lattice analysis of Ref.~\cite{Kitazawa:2019otp}.}. In this case, YM theory possesses two $Z_N$ symmetries associated with two BCs that can be spontaneously broken when $T$ and the spatial extent $L$ are varied. 
It is then expected that the nontrivial behaviors of thermodynamics observed in Ref.~\cite{Kitazawa:2019otp} on $\TTRR$ emerge in connection to the dynamics of two spontaneous symmetry breakings.
In Ref.~\cite{Suenaga:2022rjk}, this idea has been explored by extending the model in Ref.~\cite{Meisinger:2001cq} to $\mathbb{T}^2\times\mathbb{R}^2$ with two Polyakov loops. However, it was found that a simple modeling of the potential term employed in this study fails to reproduce the lattice data even qualitatively.

In the present study, we pursue this idea by improving the potential term to enhance the interplay between the Polyakov loops. In Ref.~\cite{Suenaga:2022rjk}, a simple potential term with a separable form has been employed, where the two Polyakov loops are independent in the potential term. In the present study, we introduce the cross terms describing their interaction, where the parameters in the terms are determined to reproduce the lattice data.

We show that this model succeeds in reproducing the lattice data in Ref.~\cite{Kitazawa:2019otp} qualitatively for the temperature range $T/T_{\rm d}\gtrsim1.5$. 
Moreover, we find that this model suggests the existence of a novel first-order phase transition on $\TTRR$ in the deconfined phase where both the $Z_3$ symmetries are spontaneously broken. 
This first-order transition is not connected to the deconfinement transition in the infinite volume, but terminates at the critical points on $\TTRR$ that should belong to the two-dimensional $Z_2$ universality class. 

This paper is organized as follows. In Sec.~\ref{sec:level2}, we construct the model on $\TTRR$ by extending the ideas in Refs.~\cite{Meisinger:2001cq,Dumitru:2012fw,Suenaga:2022rjk}. In Sec.~\ref{sec:NumericalResults}, we compare the thermodynamics obtained on the model with the lattice results and present the phase diagram on $\TTRR$. Physical properties of our model are discussed in more detail in Sec.~\ref{sec:physical}. 
We then give a summary and perspectives in the last section. In App.~\ref{sec:parameterdep}, the roles of parameters in the model are examined.

\section{\label{sec:level2}Effective model with two Polyakov loops}

In this section, we construct the effective model for $SU(3)$ YM theory on $\mathbb{T}^2\times\mathbb{R}^2$ with two Polyakov loops. Throughout this paper, we assume that $\tau$ and $x$ axes are compactified with the PBCs of lengths $L_\tau=1/T$ and $L_x$, respectively, while the remaining $y$ and $z$ directions are left to be infinite. 

\subsection{Polyakov loops}
\label{sec:Polyakov-loop}

Let us first briefly review the definition and properties of the Polyakov loops. 
In $SU(N)$ YM theory on $\TTRR$, one can introduce two Polyakov loops along two compactified directions 
\begin{align}
    \Omega_c(\bm{x}_c^\perp)=\frac{1}{N}{\rm Tr}\Big(\hat\Omega_c(\bm{x}_c^\perp)\Big),\qquad  (c=\tau,x), 
    \label{eq:PolyakovLoopDef}
\end{align}
where ${\rm Tr}$ denotes the trace in the $N\times N$ color space with $N$ being the number of colors and the Polyakov-loop matrix $\hat\Omega_c(\bm{x}_c^\perp)$ for $c=\tau,x$ is defined by 
\begin{align}
    \hat\Omega_c(\bm{x}_c^\perp)=\mathcal{P}\exp\Big(i\int^{L_c}_0A_c(x_c,\bm{x}_c^\perp)dx_c\Big),    
    \label{eq:PloopM}
\end{align}
with $A_\mu(x)$ being the $SU(N)$ gauge field, $\bm{x}_\tau^\perp=(x,y,z)$ and $\bm{x}_x^\perp=(\tau,y,z)$, and the path-ordering symbol $\mathcal{P}$. 

On $\TTRR$, YM theory has two center symmetries $Z^{(c)}_N$ that can be spontaneously broken. They are defined through the twisted BC for the gauge transformation, $ V(L_c,\bm{x}_c^\perp) = z_k^{(c)}V(0,\bm{x}_c^\perp)$, with $z_k^{(c)}={\rm exp}(2\pi k i/N)$ ($k=0,1\cdots, N-1$) being the center of $SU(N)$. Although the YM gauge action is invariant under the twisted gauge transformation, the Polyakov loops~\eqref{eq:PolyakovLoopDef} are not invariant~\cite{Rothe:1992nt}.
Their expectation values $\Omega_c=\langle \Omega_c(\bm{x}_c^\perp)\rangle$ thus serve as order parameters for the spontaneous symmetry breaking of the corresponding $Z^{(c)}_N$. 

For $N=3$, the Polyakov-loop matrices can be diagonalized by the gauge transformation as 
\begin{align}
    \hat\Omega_c(\bm{x}_c^\perp) = {\rm diag}[e^{(\theta_c)_1},e^{(\theta_c)_2},e^{(\theta_c)_3}],
    \label{diagP}
\end{align}
where the angle $\theta_c$'s satisfy 
\begin{align}
    \sum^3_{j=1}(\theta_c)_j=0 \qquad ({\rm mod} ~ 2\pi).
\end{align}
Using the gauge degrees of freedom, the gauge field $A_c(x)$ can be taken to be constant on $\Omega_c(\bm{x}_c^\perp)$. In this case, the gauge field corresponding to Eq.~\eqref{diagP} is given by
\begin{align}
    A_c(x_c,\bm{x}_c^\perp)=\frac{1}{L_c}{\rm diag}[(\theta_c)_1,(\theta_c)_2,(\theta_c)_3]. \label{diagA}
\end{align}

Adopting an ans\"atz~\cite{Meisinger:2001cq}
\begin{align}
    (\theta_c)_j=(2-j)\phi_c \qquad (j=1,2,3), \label{theta}
\end{align}
the corresponding Polyakov loop is expressed by a single parameter $\phi_c$ as 
\begin{align}
    \Omega_c=\frac{1}{3}\sum^3_{j=1}\exp\big(i(\theta_c)_j\big)=\frac{1}{3}(1+2\cos\phi_c). \label{Pctheta}
\end{align}
In this parametrization, the $Z_3^{(c)}$ symmetric phase with $\Omega_c=0$ is realized at $\phi_c=2\pi/3$, while $\phi_c=0$ corresponds to $\Omega_c=1$.
In the following, we assume 
$0\leq\phi_c\leq2\pi/3$.

\subsection{Model construction}

To describe the thermodynamics of $SU(3)$ YM theory on $\mathbb{T}^2\times\mathbb{R}^2$, we employ an effective model with two Polyakov loops $\Omega_\tau(\bm{x}_\tau^\perp)$ and $\Omega_x(\bm{x}_x^\perp)$~\cite{Suenaga:2022rjk}. This model has been introduced as an extension of Ref.~\cite{Meisinger:2001cq}, where the thermodynamics in an infinitely large volume, i.e. $\SRRR$, has been investigated by an effective model with a single Polyakov loop $\Omega_\tau(\bm{x}_\tau^\perp)$ that is relevant in this case. In Ref.~\cite{Meisinger:2001cq}, it is assumed that $\Omega_\tau(\bm{x}_\tau^\perp)$ is spatially uniform and its expectation value $\Omega_\tau$ is determined so as to minimize the free-energy density consisting of two parts 
\begin{align}
    f^{\SRRR}(\vec\theta_\tau;L_\tau)
    =f_{\rm pert}^{\SRRR}(\vec\theta_\tau;L_\tau)+f_{\rm pot}^{\SRRR}(\vec\theta_\tau;L_\tau),
    \label{ftot}
\end{align}
with $\vec\theta_c=((\theta_c)_1,(\theta_c)_2,(\theta_c)_3)$. Here, $f_{\rm pert}^{\SRRR}(\vec\theta_\tau;L_\tau)$ is the contribution from perturbative gluons upon the background gauge field Eq.~\eqref{diagA}, while $f_{\rm pot}^{\SRRR}(\vec\theta_\tau;L_\tau)$ represents the phenomenological potential term describing nonperturbative effects leading to the confined phase at low $T$. 
In Ref.~\cite{Meisinger:2001cq}, two forms of $f_{\rm pot}^{\SRRR}(\vec\theta_\tau;L_\tau)$ have been employed and it was found that in both modelings the energy density $\epsilon$ and pressure $p$ calculated from Eq.~\eqref{ftot} can qualitatively reproduce the lattice results~\cite{Boyd:1996bx}, especially the characteristic behavior of the interaction measure $\Delta=\varepsilon-3p$ near $T_{\rm d}$.

This idea has been elaborated later in Ref.~\cite{Dumitru:2012fw}, where the following form of the potential term is employed~\footnote{
We employ the ``two-parameter model'' in Ref.~\cite{Dumitru:2012fw}, which has the best agreement with the lattice data.
The last term in Eq.~(\ref{FDumitru}) is divided by two from the original one so that the constant term in the separable potential on $\TTRR$ in Eq.~\eqref{eq:fsep} agrees with the original one.}: 
\begin{align}
f^{\SRRR}_{\rm pot}(\vec{\theta},L) 
=&-\frac{4\pi^2}{3}\frac{T_{\rm d}^2}{L^2}\Big(\frac{1}{5}c_1 V_1(\phi)+c_2 V_2(\phi)-\frac{2}{15}c_3\Big) \notag \\
&+\frac{\tilde c_3}2\frac{8\pi^2T^4_{\rm d}}{45} , 
\label{FDumitru}
\\
V_1(\phi)=\,& \frac{2\phi(\pi-\phi)+\phi(2\pi-\phi)}{2\pi^2} ,
\label{eq:V1}
\\
V_2(\phi)=\,& \frac{8\phi^2(\pi-\phi)^2+\phi^2(2\pi-\phi)^2}{8\pi^4},
\label{eq:V2}
\end{align}
for $0\leq \phi\leq 2\pi/3$~\footnote{The periodicities with respect to $\phi\to\phi+2\pi$ and $\phi\to\phi+\pi$ are apparently lost in Eqs.~(\ref{eq:V1}) and~(\ref{eq:V2}), since we have rewritten the periodic Bernoulli polynomials by the corresponding ordinary ones so as to express the potential by elementary functions, which is justified within the range $0\leq \phi\leq 2\pi/3$. For more detail see Ref.~\cite{Dumitru:2012fw}.}
with $\vec\theta=(\phi,0,-\phi)$ and $\tilde c_3=(47-20c_2-27c_3)/27$, where the parameters are determined to reproduce the lattice data as 
\begin{align}
    c_1=0.552, \ c_2=0.830, \ c_3=0.950.
    \label{RobParameter}
\end{align}
Since this model has better agreement with the lattice results than those in Ref.~\cite{Meisinger:2001cq}, in the present study we employ Eq.~\eqref{FDumitru} as the limiting form of the potential term in our model on $\TTRR$.

We note that thermodynamics calculated with Eq.~\eqref{FDumitru} has at most about $10\%$ deviation from the lattice data. This deviation is inherited to our model as we will see in the next section. We regard this deviation 
negligible for our study that investigates 
qualitative roles of two Polyakov loops on $\TTRR$.

By extending the models in Refs.~\cite{Meisinger:2001cq,Dumitru:2012fw} to $\TTRR$, we assume that the free-energy density contains two Polyakov loops $\Omega_\tau$ and $\Omega_x$ and is given by two terms as~\cite{Suenaga:2022rjk}
\begin{align}
    &f(\vec\theta_\tau,\vec\theta_x;L_\tau,L_x)
    \notag \\
    &= f_{\rm pert}(\vec\theta_\tau,\vec\theta_x;L_\tau,L_x)
    + f_{\rm pot}(\vec\theta_\tau,\vec\theta_x;L_\tau,L_x),
    \label{eq:f}
\end{align}
which is now a function of $L_\tau$ and $L_x$. The forms of $f_{\rm pert}$ and $f_{\rm pot}$ will be specified below.
The values of $\Omega_\tau$ and $\Omega_x$ are determined to minimize Eq.~\eqref{eq:f}.

\subsection{Perturbative contributions}
\label{sec:FPert}

Let us first specify the perturbative term $f_{\rm pert}(\vec\theta_\tau,\vec\theta_x;L_\tau,L_x)$.
On $\TTRR$, employing the massless gluons and assuming the simultaneous diagonalizations of $\Omega_\tau$ and $\Omega_x$, the free-energy density of perturbative gluons on the background gauge field $A_\tau$ and $A_x$ in Eq.~\eqref{diagA} is given by~\cite{Sasaki:2013xfa,Suenaga:2022rjk}
\begin{align}
&f_{\rm pert}(\vec{\theta}_\tau,\vec{\theta}_x;L_\tau,L_x) \notag \\
&=\frac{1}{L_\tau L_x}\sum^3_{j,k=1}\Big(1-\frac{\delta_{jk}}{3}\Big)\sum_{\ell_\tau,\ell_x}\int \frac{d^2p_L}{(2\pi)^2}\ln\Big[\Big(\omega_\tau \nonumber\\
& -\frac{(\Delta\theta_\tau)_{jk}}{L_\tau}\Big)^2 +\Big(\omega_x-\frac{(\Delta\theta_x)_{jk}}{L_x}\Big)^2+\bm{p}^2_L\Big], \label{fpert1}
\end{align}
with the generalized Matsubara modes $\omega_c=2\pi \ell_c/L_c$ for integer $\ell_c$, the transverse momenta $\bm{p}_L=(p_y,p_z)$, and $(\Delta\theta_c)_{jk}=(\theta_c)_j-(\theta_c)_k$. 

Equation~(\ref{fpert1}) contains the double infinite summations for $\ell_\tau$ and $\ell_x$, which give rise to ultraviolet (UV) divergences. Following the procedure in Ref.~\cite{Suenaga:2022rjk},
the summation is rearranged with the aid of the generalized Epstein-Hurwitz zeta function with the UV subtraction. The resultant form reads 
\begin{align}
&f_{\rm pert}(\vec{\theta}_\tau,\vec{\theta}_x;L_\tau,L_x) \notag \\
&= -\frac{8\pi^2}{45L^4_\tau}+\frac{8\phi^2_\tau(\phi_\tau-\pi)^2+\phi^2_\tau(\phi_\tau-2\pi)^2}{6\pi^2L^4_\tau} \notag \\
&-\frac{8\pi^2}{45L^4_x}+\frac{8\phi^2_x(\phi_x-\pi)^2+\phi^2_x(\phi_x-2\pi)^2}{6\pi^2L^4_x} \notag \\
&-\frac{8}{\pi^2}\sum^\infty_{l_\tau,l_x=1}\frac{1}{{X^4_{l_\tau,l_x}}}\Big[1+2\cos(\phi_\tau l_\tau)\cos(\phi_x l_x) \notag \\
&+\cos(2\phi_\tau l_\tau)\cos(2\phi_x l_x)\Big],\label{fpert2}
\end{align}
with $\vec\theta_c=(\phi_c,0,-\phi_c)$ and
\begin{align}
X_{l_\tau,l_x}\equiv\sqrt{(l_\tau L_\tau)^2+(l_xL_x)^2}.
\end{align}

As shown in App.~\ref{doublesum},
the double summation in Eq.~\eqref{fpert2} is further reduced to a single-sum form as 
\begin{align}
&\sum^\infty_{l_\tau,l_x=1}\frac{1}{{X^4_{l_\tau,l_x}}}\Big[1+2\cos(\phi_\tau l_\tau)\cos(\phi_x l_x) 
    \notag \\
    &\hspace{2cm} 
    +\cos(2\phi_\tau l_\tau)\cos(2\phi_x l_x)\Big] 
    \notag \\ &=\sum^\infty_{l_x=1}\frac{1}{L_\tau^4}\Big[G\Big( 0,l_x \frac{L_x}{L_\tau} \Big)+2\cos(\phi_x l_x)G\Big(\phi_\tau,l_x \frac{L_x}{L_\tau} \Big)
    \notag \\
    &\hspace{2cm} 
    +\cos(2\phi_x l_x)G\Big(2\phi_\tau,l_x \frac{L_x}{L_\tau}\Big)\Big], \label{doublesumconcrete}
\end{align}
with
\begin{align}
    G(a,C)=& \, \frac{\pi}{4C^2}\frac{1}{\sinh(\pi C)}\left[\frac{\cosh[(\pi-a)C]}{C}\right.\notag \\
    &\left.+a \sinh[(\pi-a)C]+\pi\frac{\cosh(a C)}{\sinh(\pi C)}\right]-\frac{1}{2C^4}. 
    \label{eq:g(a,C)}
\end{align}
The use of Eq.~\eqref{doublesumconcrete} reduces the numerical cost for the calculation of $f_{\rm pert}$ drastically.

In the limit $L_x\to\infty$, only the first two terms in Eq.~\eqref{fpert2} survive and Eq.~\eqref{fpert2} reduces to $f_{\rm pert}^{\SRRR}$~\cite{Meisinger:2001cq}. We note that the sub-leading terms with respect to $L_\tau/L_x$ in this limit start at order $(L_\tau/L_x)^3$~\cite{Suenaga:2022rjk}:
\begin{align}
    f_{\rm pert}(\vec\theta_\tau,\vec\theta_x;L_\tau,L_x) 
    = f_{\rm pert}^{\SRRR}(\vec\theta_\tau;L_\tau) + {\cal O}(L_x^{-3}),
    \label{eq:fpertLx->inf}
\end{align}
where the ${\cal O}(L_x^{-3})$ term is nonzero only when $\phi_\tau=0$. Eq.~\eqref{eq:fpertLx->inf} can be confirmed easily using Eqs.~\eqref{doublesumconcrete} and~\eqref{eq:g(a,C)}. 

\subsection{\label{sec:citeref}Potential term}

Next, let us consider the potential term $f_{\rm pot}$ in Eq.~\eqref{eq:f}. Since the minimum of Eq.~\eqref{fpert2} is at $\Omega_\tau=\Omega_x=1$ and it solely leads to the $Z_3^{(c)}$ broken phase, the potential term is necessary to realize the symmetric phase at large $L_\tau$ and $L_x$.

Because we introduce this term in a phenomenological manner, we first list the general constraints on its form:
\begin{enumerate}
    \item[(i)] 
    The free energy~\eqref{eq:f} should be invariant under the exchange of $\tau$ and $x$ axes, $f(\vec{\theta}_\tau,\vec{\theta}_x;L_\tau,L_x)=f(\vec{\theta}_x,\vec{\theta}_\tau;L_x,L_\tau)$. As $f_{\rm pert}$ in Eq.~\eqref{fpert1} is in agreement with this condition, $f_{\rm pot}$ must satisfy 
    \begin{align}
        f_{\rm pot}(\vec{\theta}_\tau,\vec{\theta}_x;L_\tau,L_x)=f_{\rm pot}(\vec{\theta}_x,\vec{\theta}_\tau;L_x,L_\tau).
        \label{eq:f_tx}
    \end{align}
    
    \item[(ii)] 
    In the $L_x\rightarrow\infty$ limit, the system reduces to $\SRRR$ where the conventional thermodynamics must be recovered.
    This requirement is satisfied if $f_{\rm pot}$ approaches the one on $\SRRR$,
    \begin{align}
        f_{\rm pot}(\vec{\theta}_\tau,\vec{\theta}_x;L_\tau,L_x)
        \xrightarrow[L_x\rightarrow\infty]{} 
        f_{\rm pot}^{\mathbb{S}^1\times\mathbb{R}^3}(\vec{\theta}_\tau,L_\tau).
        \label{reducepot}
    \end{align}
    In this study we employ Eq.~\eqref{FDumitru} for $f_{\rm pot}^{\mathbb{S}^1\times\mathbb{R}^3}$. From Eq.~\eqref{eq:f_tx}, we also obtain
    \begin{align}
        f_{\rm pot}(\vec{\theta}_\tau,\vec{\theta}_x;L_\tau,L_x)
        \xrightarrow[L_\tau\rightarrow\infty]{}
        f_{\rm pot}^{\mathbb{S}^1\times\mathbb{R}^3}(\vec{\theta}_x,L_x). 
        \label{reducepot2}
    \end{align}

    \item[(iii)] 
    In the $L_\tau\rightarrow\infty$ limit, the system should be in the confined phase irrespective of the value of $L_x$, which means that 
    \begin{align}
        \Omega_\tau \xrightarrow[L_\tau\rightarrow\infty]{} 0  \qquad ({\rm for \ any} \ L_x),
        \label{eq:Ptau0}
        \\
        \Omega_x \xrightarrow[L_x\rightarrow\infty]{} 0  \qquad ({\rm for \ any} \ L_\tau),        \label{eq:Px0}
    \end{align}
    where Eq.~\eqref{eq:Px0} is from the constraint~(i).
    To realize Eq.~\eqref{eq:Px0}, in the $L_x\to\infty$ limit $\vec\theta_x$ dependence of $f_{\rm pot}$ must dominate over $f_{\rm pert}$, and it leads to Eq.~\eqref{eq:Px0}. From Eq.~\eqref{eq:fpertLx->inf}, this means that $f_{\rm pot}(\vec{\theta}_\tau,\vec{\theta}_x;L_\tau,L_x)$ must have $\vec\theta_x$ dependence that falls off slower than $1/L_x^3$ for $L_x\to\infty$. The same argument also applies to the $L_\tau\to\infty$ limit.

    \item[(iv)] 
    In the limit $L_\tau\to0$ and $L_x\to0$, the perturbative term $f_{\rm pert}$ must dominate over $f_{\rm pot}$ so that the system approaches the free gluon gas in this limit. Also, in this limit both the $Z_3^{(\tau)}$ and $Z_3^{(x)}$ symmetries are explicitly broken with
    \begin{align}
        \Omega_\tau=\Omega_x=1 \qquad (L_x,L_\tau\rightarrow 0) .
    \end{align}
\end{enumerate}

In Ref.~\cite{Suenaga:2022rjk}, as a potential term satisfying these constraints, the ``separable'' form 
\begin{align}
    f_{\rm sep}(\vec{\theta}_\tau,\vec{\theta}_x;L_\tau,L_x)=
    f^{\SRRR}_{\rm pot}(\vec{\theta}_\tau,L_\tau)
    +f^{\SRRR}_{\rm pot}(\vec{\theta}_x,L_x),
    \label{eq:fsep}
\end{align}
has been employed for $f_{\rm pot}(\vec{\theta}_\tau,\vec{\theta}_x;L_\tau,L_x)$ as a simple extension of Ref.~\cite{Meisinger:2001cq}. However, it has been found that this potential term cannot reproduce the lattice results on $\TTRR$ even qualitatively. While the ``model-B'' in Ref.~\cite{Meisinger:2001cq} has been used for $f^{\SRRR}_{\rm pot}(\vec{\theta}_x,L_x)$ in Ref.~\cite{Suenaga:2022rjk}, we have checked that the result hardly changes even if we employ other potential terms proposed in Refs.~\cite{Meisinger:2001cq,Dumitru:2012fw}. This result implies that $f_{\rm pot}$ should contain cross terms of $\Omega_\tau$ and $\Omega_x$ that physically describe their interplay.

To introduce such effects into $f_{\rm pot}$ keeping the constraints (i)--(iv), in the present study we assume that the potential term is given by
\begin{align}
    f_{\rm pot}(\vec{\theta}_\tau,\vec{\theta}_x;L_\tau,L_x)
    = \, &
    f_{\rm sep}(\vec{\theta}_\tau,\vec{\theta}_x;L_\tau,L_x)
    \notag \\
    &+f_{\rm cross}(\vec{\theta}_\tau,\vec{\theta}_x;L_\tau,L_x), \label{FPotSeparate}
\end{align}
i.e. we add the cross term $f_{\rm cross}$ on top of $f_{\rm sep}$.

Constraints on $f_{\rm cross}$ are in order: 
\begin{enumerate}
\item[(v)] 
To satisfy the conditions~\eqref{reducepot} and~\eqref{reducepot2}, $f_{\rm cross}$ must drop faster than $f_{\rm sep}(\vec{\theta}_\tau,\vec{\theta}_x;L_\tau,L_x)$ in the limit $L_\tau\to\infty$ or $L_x\to\infty$:
\begin{align}
    \frac{f_{\rm cross}(\vec{\theta}_\tau,\vec{\theta}_x;L_\tau,L_x)}{f_{\rm sep}(\vec{\theta}_\tau,\vec{\theta}_x;L_\tau,L_x)} \xrightarrow[L_\tau,~L_x\to\infty]{} 0.
    \label{cross/sep->0}
\end{align}

\item[(vi)]
To satisfy the constraint~(iii), $\vec\theta_x$ dependence in $f_{\rm cross}$ must drop faster than that in $f_{\rm sep}$ in the limit $L_x\to\infty$, which is ${\cal O}(L_x^{-2})$ from Eq.~\eqref{eq:fsep}~\footnote{This condition can be relaxed if $f_{\rm cross}$ solely leads to Eq.~\eqref{eq:Px0} in this limit. As we will see later, however, Eq.~\eqref{eq:fcross} with our parameter choice~\eqref{FCrossParameter} does not satisfy it. We thus require this constraint in this study.}.

\item[(vii)]
Equation~\eqref{eq:f_tx} requires
\begin{align}
    f_{\rm cross}(\vec{\theta}_\tau,\vec{\theta}_x;L_\tau,L_x) = f_{\rm cross}(\vec{\theta}_x,\vec{\theta}_\tau;L_x,L_\tau).
\end{align}

\item [(viii)]
Terms in $f_{\rm cross}$ should be real and invariant under the $Z_3^{(c)}$ transformations. 
$f_{\rm cross}$ thus depends on $\Omega_c$ only through the $Z_3^{(c)}$-invariant combinations
\begin{align}
    X_c^{(1)} =&\, {\rm Tr}(\hat \Omega_c) {\rm Tr}(\hat \Omega_c^\dagger) , 
    \label{eq:Omega1} \\
    X_c^{(2)} =&\, \frac12 \big[ {\rm Tr}(\hat \Omega_c^2) {\rm Tr}(\hat \Omega_c) + {\rm Tr}(\hat \Omega_c^{\dagger2}) {\rm Tr}(\hat \Omega_c^\dagger) \big], 
    \label{eq:Omega2} \\
    X_c^{(3)} =&\, \frac12 \big[ {\rm Tr}(\hat \Omega_c^3) + {\rm Tr}(\hat \Omega_c^{\dagger3}) \big], 
    \label{eq:Omega3} \\
    X_c^{(4)} =&\, \frac12 \big[ ({\rm Tr}(\hat \Omega_c))^3 + ({\rm Tr}(\hat \Omega_c^\dagger))^3 \big] ,
    \label{eq:Omega4} 
\end{align}
up to third order in $\Omega_c$. Although yet higher order terms are allowed from the symmetry, in this study we assume that $f_{\rm cross}$ does not contain them to keep simplicity.
Among Eqs.~\eqref{eq:Omega1}--\eqref{eq:Omega4}, only two terms are independent in the sense that the others can be written by their linear combinations. In fact, by choosing $X_c^{(1)}$ and $X_c^{(3)}$ as independent terms, Eqs.~\eqref{eq:Omega2} and~\eqref{eq:Omega4} are written as
\begin{align}
    X_c^{(2)} = \, & X_c^{(1)}+ X_c^{(3)}, \\
    X_c^{(4)} = \, & 3X_c^{(1)}+X_c^{(3)}-6.
\end{align}
\end{enumerate}

For the functional form of $f_{\rm cross}$ satisfying these constraints, in the present study we employ 
\begin{align}
    &f_{\rm cross}(\vec{\theta}_\tau,\vec{\theta}_x;L_\tau,L_x)
    \notag \\
    &= g(L_\tau,L_x) \Big[ c_4 X_\tau^{(1)}X_x^{(1)}
    + c_5 ( X_\tau^{(3)}X_x^{(1)} + X_\tau^{(1)}X_x^{(3)} )
    \notag \\
    & \hspace{22mm} + c_6 X_\tau^{(3)}X_x^{(3)} \Big],
    \label{eq:fcross}
\end{align}
where $X_c^{(1)}$ and $X_c^{(3)}$ are used for independent terms in Eqs.~\eqref{eq:Omega1}--\eqref{eq:Omega4} and the coefficients $c_4, c_5, c_6$ are dimensionless parameters to be determined to reproduce the lattice data. $g(L_\tau,L_x)=g(L_x,L_\tau)$ controls $L_\tau$ and $L_x$ dependence. We assume a simple ans\"atz
\begin{align}
    g(L_\tau,L_x) = \frac{T_{\rm d}^4}{\big[ T_{\rm d}^2(L_\tau^2+L_x^2)\big]^n}, 
    \label{fTypeI} 
\end{align}
where $T_{\rm d}$ is used to make this term dimensionless and $n$ is the other dimensionless free parameter in our model. 
With the parametrization~\eqref{theta}, $X_c^{(1)}$ and $X_c^{(3)}$ are given as functions of $\phi_c$ as
\begin{align}
    X_c^{(1)} &= ( 1 + 2 \cos\phi_c)^2 ,
    \label{eq:Omega1phi}
    \\
    X_c^{(3)} &= 1 + 2 \cos(3\phi_c) .
    \label{eq:Omega3phi}
\end{align}

The value of $n$ is limited by the constraints~(iv) and~(vi). 
First, from~(iv), $f_{\rm cross}$ needs to be suppressed faster than $f_{\rm pert}$ in the limit $L_\tau,L_x\to0$ with fixed $L_\tau/L_x$. This gives the upper bound $n<2$. Second, from~(vi) $f_{\rm cross}$ should drop faster than $L_x^{-2}$ in the $L_x\to\infty$ limit. This requires $n>1$. Therefore, $n$ must satisfy
\begin{align}
1<n<2. \label{NConstraint}
\end{align}
This completes the model construction. Assumptions and ans\"atze of our model are summarized in App.~\ref{ansatz}.

Before closing this subsection, we give several comments on properties and assumptions imposed in the above model construction. 
First, our model assumes Eq.~\eqref{FPotSeparate}. We use this form to fulfill the constraints~(i) and (ii) in a simple manner, although another parametrization is also possible. Second, we truncate the terms in $f_{\rm cross}$ at the third order in $\Omega_c$. However, there are no a priori reasons to justify this assumption because $\Omega_c$ is not small in general on $\TTRR$. It, however, is worth emphasizing that Eq.~\eqref{eq:fcross} contains $X_c^{(3)}$ that is specific to $SU(3)$ gauge theory. 
Third, while we attribute $L_\tau$ and $L_x$ dependence to an overall factor $g(L_\tau,L_x)$, each term in $f_{\rm cross}$ can have different dependence on $L_\tau$ and $L_x$ in general. We, however, do not consider this possibility just to suppress the number of 
free parameters in the model.

In spite of the simple modeling, as shown below our model is capable of reproducing the lattice data of thermodynamics on ${\mathbb T}^2\times {\mathbb R}^2$ qualitatively for $T/T_{\rm d}\gtrsim1.5$.

\subsection{Thermodynamics}
\label{sec:model:therm}

When BCs are imposed in thermal systems, pressure is no longer necessarily isotropic since the rotational symmetry is broken by the BCs. In our setup where the PBC is imposed in $x$ direction, the pressure along $x$ direction, $p_x$, can be different from those along $y$ and $z$ directions $p_y$ and $p_z$, while $p_y=p_z$ holds owing to the rotational symmetry in the $y$--$z$ plane. The energy-momentum tensor $T^\mu_\nu(x)$ in this situation is given by
\begin{align}
    T^\mu_\nu(x)={\rm diag}(\epsilon,p_x,p_z,p_z) ,
    \label{EMT}
\end{align}
with the energy density $\epsilon$, where $T^\mu_\nu(x)$ is diagonalized in this system owing to reflection symmetries along $x,y,z$ directions.
For $L_\tau=L_x$, the $\tau$ and $x$ directions in the Euclidean spacetime are degenerated and this yields~\cite{Kitazawa:2019otp}
\begin{align}
    p_x=-\epsilon \qquad {\rm at}~ L_\tau=L_x.
    \label{eq:px=e}
\end{align}

Thermodynamic quantities $\epsilon,p_x,p_y,p_z$ on $\TTRR$ are evaluated from the free-energy density $f$ in Eq.~\eqref{eq:f} as 
\begin{align}
    \epsilon=& \, \frac{L_\tau}{\mathcal{V}}\frac{\partial}{\partial L_\tau}\Big(\mathcal{V}f\Big) , 
    \label{Energy} \\ 
    p_x=&-\frac{L_x}{\mathcal{V}}\frac{\partial}{\partial L_x}\Big(\mathcal{V}f\Big)  ,
    \label{PressureX} \\ 
    p_y = p_z=&-\frac{L_z}{\mathcal{V}}\frac{\partial}{\partial L_z}\Big(\mathcal{V}f\Big) = -f, 
    \label{PressureZ}
\end{align}
with $\mathcal{V}=L_\tau L_xL_yL_z$. In the last equality of Eq.~\eqref{PressureZ} we used the fact that $f$ does not depend on $L_z$.
When the system possesses the scale invariance, the interaction measure 
\begin{align}
    \Delta = T^\mu_\mu = \epsilon-p_x-2p_z ,
\end{align}
vanishes. In this case, we have $p_x/p_z=-1$ at $L_\tau=L_x$ from Eq.~\eqref{eq:px=e}.

\subsection{Parameters}

There are four free paramters in our model; $c_4, c_5, c_6, n$. They are determined to reproduce the lattice data on thermodynamics in Ref.~\cite{Kitazawa:2019otp}. After surveying their parameter dependence on $\TTRR$, we have found that the parameter set 
\begin{align}
    c_4=0.11, \ c_5=0.06, \ c_6=-0.03, \ n=1.85, \label{FCrossParameter}
\end{align}
gives a reasonable agreement with the lattice data over a wide range of $L_\tau$ and $L_x$. We thus employ Eq.~\eqref{FCrossParameter} in what follows.
We, however, also found that our model with Eq.~\eqref{FCrossParameter} reproduces the lattice data only for $T/T_{\rm d}\gtrsim1.5$ as we will see in Sec.~\ref{sec:ThermodynamicsLattice}, while the lattice data in Ref.~\cite{Kitazawa:2019otp} are available for $1.12\le T/T_{\rm d}\le 25$. As far as we have checked, no parameter set can well reproduce all the lattice data. The dependence of thermodynamics on each parameter is discussed in App.~\ref{sec:parameterdep}.

\begin{figure}
\includegraphics[width=0.48\textwidth]{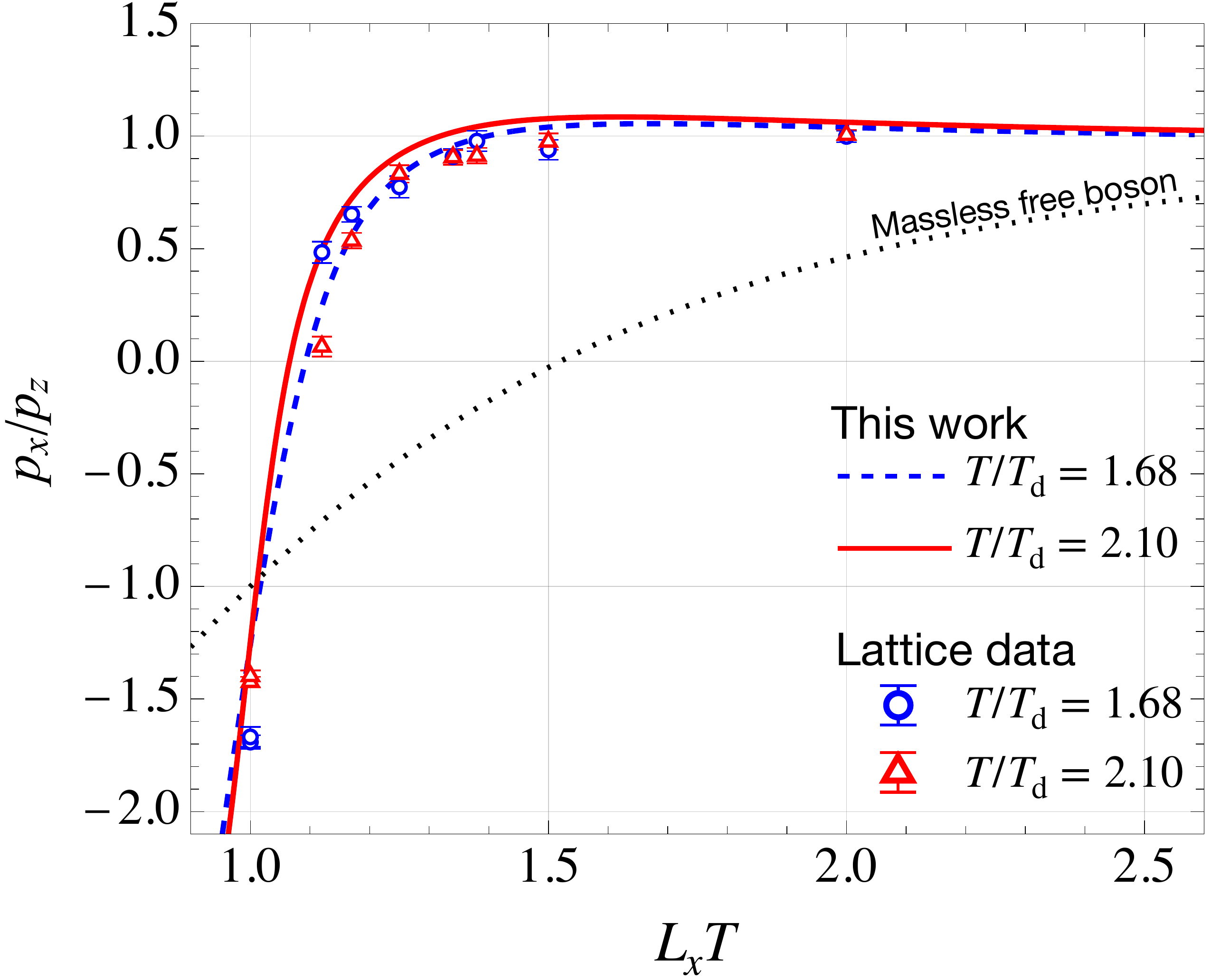}
\caption{\label{pxpz1} 
$L_xT$ dependence of the ratio $p_x/p_z$ at $T/T_{\rm d}=1.68$ and $2.10$, together with the lattice data~\cite{Kitazawa:2019otp}. The dotted line shows the ratio in the massless free-boson system.}
\end{figure}

\begin{figure}
\includegraphics[width=0.45\textwidth]{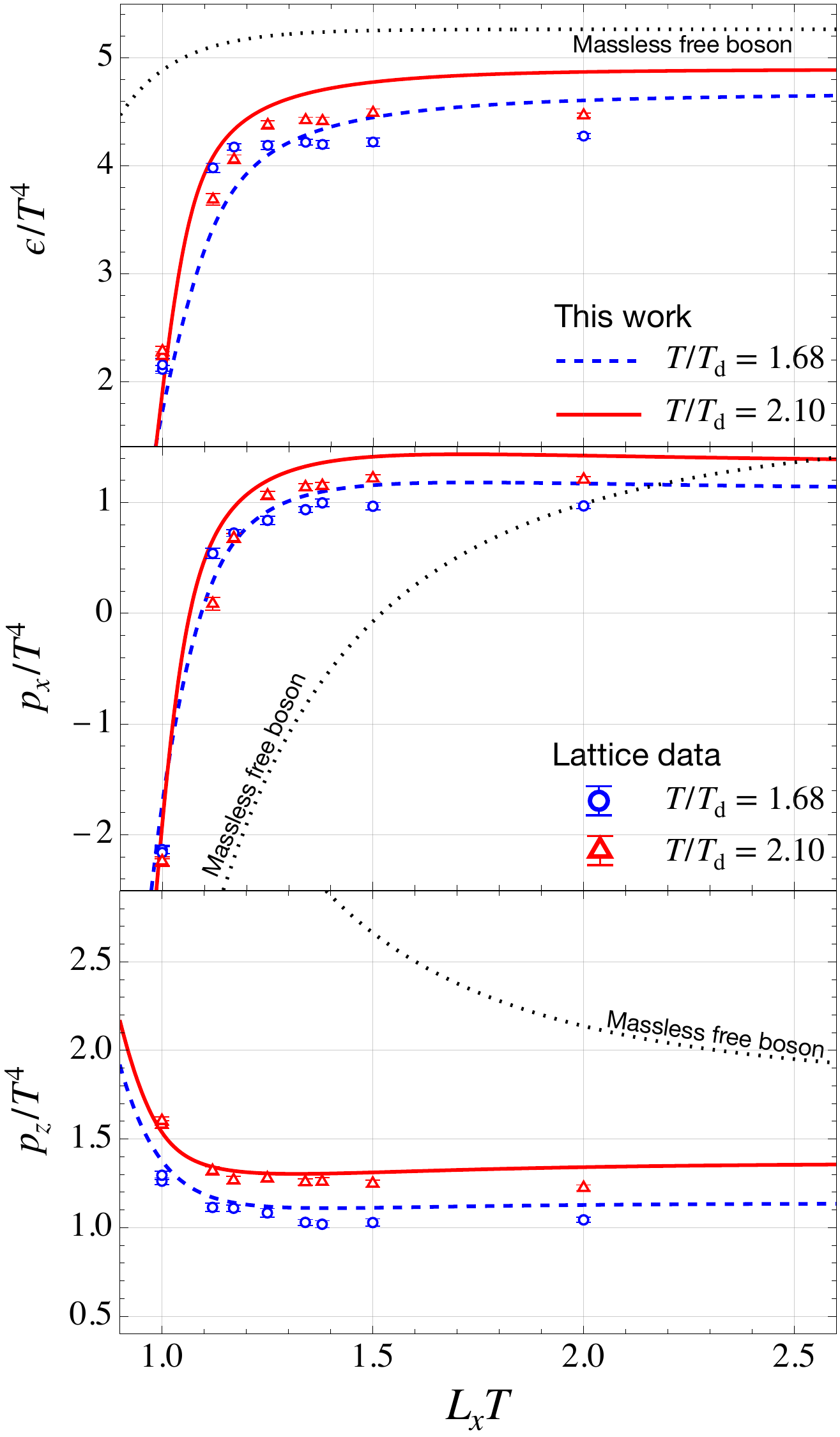}
\caption{\label{epxpz} 
$L_xT$ dependence of the energy density $\epsilon$ (top) and the pressures $p_x,\ p_z$ (middle and bottom) at $T/T_{\rm d}=1.68$ and $2.10$. The lattice data in Ref.~\cite{Kitazawa:2019otp} are also indicated by the discrete points.}
\end{figure}

\section{Numerical Results}
\label{sec:NumericalResults}

\subsection{Thermodynamic quantities}
\label{sec:ThermodynamicsLattice}

\begin{figure*}[t]
    \centering
    \includegraphics*[width=1\textwidth]{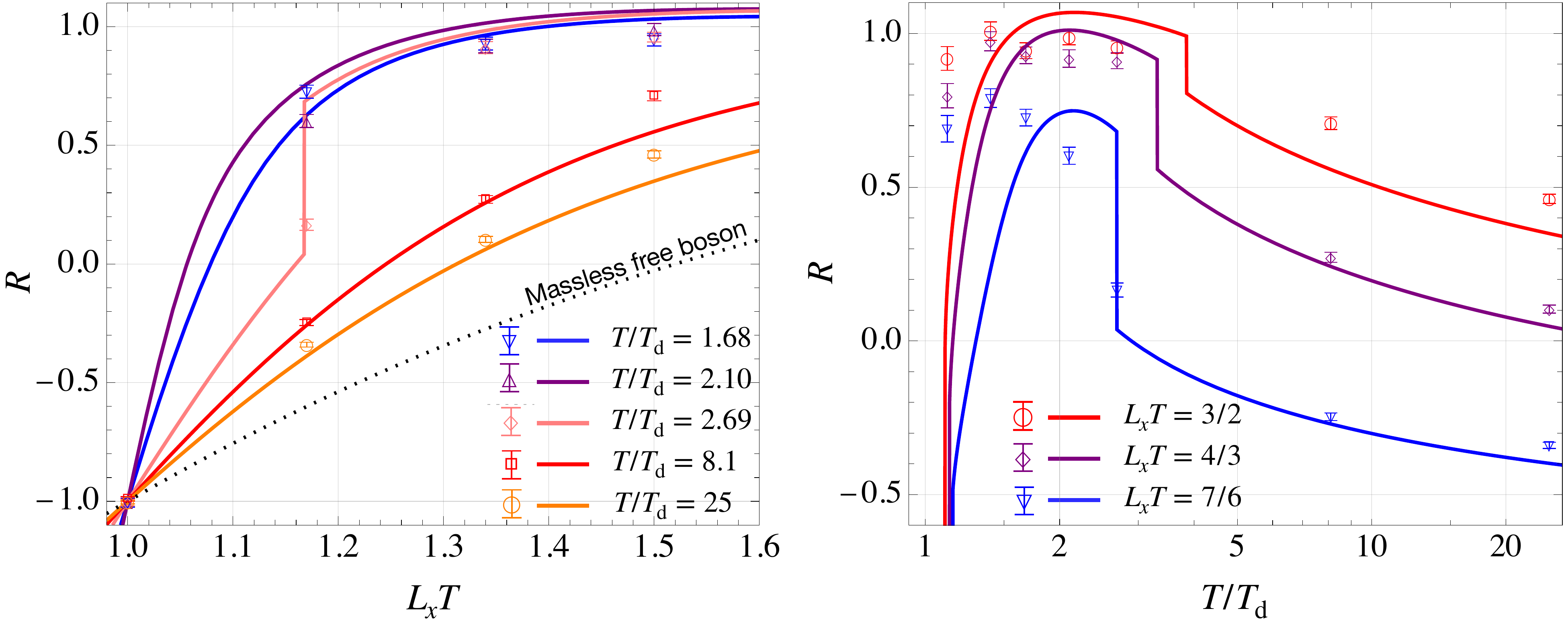}
 \caption{
 Left: $L_xT$ dependence of the ratio $R=(p_x+\Delta/4)/(p_z+\Delta/4)$ in Eq.~\eqref{ratio} for several values of $T/T_{\rm d}$. Right: $T/T_{\rm d}$ dependence of $R$ for various $L_xT$. }
 \label{HighTRatio} 
\end{figure*}

Now, let us compare the behaviors of thermodynamic quantities~(\ref{Energy})--(\ref{PressureZ}) in the model constructed above with the lattice data on ${\mathbb T}^2\times{\mathbb R}^2$.

In Fig.~\ref{pxpz1}, we first show the $L_x T$ dependence of the ratio $p_x/p_z$ at $T/T_{\rm d}=1.68$ (blue-dashed) and $2.10$ (red-solid) together with the lattice data in Ref.~\cite{Kitazawa:2019otp} indicated by discrete points with errorbars. Since $p_x/p_z=1$ in an isotropic system, the ratio satisfies $\lim_{L_xT\to\infty}p_x/p_z=1$ and its deviation from unity is a measure of anisotropy. In the figure, the behavior of $p_x/p_z$ in the massless free-boson system is also shown by the dotted line. Whereas the dotted line has a significant deviation from unity already at $L_xT=2.5$, the lattice data stay around $p_x/p_z=1$ even at $L_xT=1.3$ and then suddenly drops to a negative value at $L_xT\lesssim1.2$. 

The figure shows that these lattice results are qualitatively reproduced by our model. This result is contrasted to that in Ref.~\cite{Suenaga:2022rjk} obtained without the cross term. This difference indicates that the interplay between $\Omega_\tau$ and $\Omega_x$ induced by the cross term plays an important role on $\TTRR$. We will discuss the roles of the cross term in more detail in Sec.~\ref{sec:physical}.

Next, in Fig.~\ref{epxpz} we show the $L_xT$ dependence of $\epsilon,\ p_x$, and $p_z$ at $T/T_{\rm d}=1.68$ and $2.10$. The figure shows that the model results shown by the lines qualitatively reproduce the lattice data, while the model results slightly depart from the lattice data at large $L_xT$. These deviations are mainly attributed to the use of Eq.~\eqref{FDumitru} for $f_{\rm pot}^{\SRRR}$. As discussed already, in the large $L_xT$ limit thermodynamics in our model converges to those in Ref.~\cite{Dumitru:2012fw}, where the lattice data are not accurately reproduced. This deviation is carried over to our study.
We, however, do not modify $f_{\rm pot}^{\SRRR}$ to concentrate on the qualitative roles of $\Omega_\tau$ and $\Omega_x$ on $\TTRR$.

In lattice numerical simulations, the analysis of thermodynamics becomes more difficult for high temperatures. 
In Ref.~\cite{Kitazawa:2019otp}, as a thermodynamic quantity that can be analyzed avoiding technical problems, the ratio 
\begin{align}
 R = \frac{p_x+\Delta/4}{p_z+\Delta/4}, 
 \label{ratio}
\end{align}
has been investigated for high $T$.
In Fig.~\ref{HighTRatio}, we compare the model results with the lattice data in terms of Eq.~\eqref{ratio} for temperatures up to $T/T_{\rm d}\simeq25$. In the left panel, $R$ is plotted as functions of $L_xT$ for various $T/T_{\rm d}$, while the right panel shows the same quantity as functions of $T/T_{\rm d}$ for several values of $L_xT$. We note that Eq.~\eqref{eq:px=e} gives $R=-1$ at $L_xT=1$, which is satisfied in both the model and lattice results.

From the left panel of Fig.~\ref{HighTRatio}, one sees that the lattice data of $R$ behave differently for low and high temperatures. At $T/T_{\rm d}=1.68$ and $2.10$, $R$ changes drastically at $L_xT\simeq1.2$ as is consistent with Fig.~\ref{pxpz1}. On the other hand, results at $T/T_{\rm d}=8.1$ and $25$ behave more smoothly. In between, at $T/T_{\rm d}=2.69$ the behavior for $L_xT\gtrsim1.3$ is consistent with the lower-$T$ ones, but a data point at $L_xT=7/6\simeq1.17$ has a drop.

These lattice results are nicely reproduced by the model as shown by the lines. Remarkably, the model result at $T/T_{\rm d}=2.69$ has a discontinuous jump, i.e. a first-order phase transition, at $L_xT\simeq 1.17$.

The first-order phase transition is more clearly seen in the right panel of Fig.~\ref{HighTRatio}, where the model gives a discontinuous jump of $R$ for each $L_xT$. 
Unfortunately, the lattice data in Ref.~\cite{Kitazawa:2019otp} are too coarse to verify the existence of 
the discontinuity. However, the lattice data at $L_xT=4/3,~7/6$ have a rapid change around $T/T_{\rm d}\simeq3$, which indeed implies the existence of the first-order transition.

The right panel of Fig.~\ref{HighTRatio} also shows that our model fails in reproducing the lattice data for $T/T_{\rm d}\lesssim1.5$. In particular, in the model result $R$ drops toward negative values as $T/T_{\rm d}$ is lowered toward unity while the lattice data in Ref.~\cite{Kitazawa:2019otp} do not have such behaviors. Our model thus is not well applicable to this $T$ range. We also note that our model does not reproduce the lattice data quantitatively even for $T/T_{\rm d}\gtrsim1.5$.
For example, the left panel of Fig.~\ref{HighTRatio} shows that the temperature dependence is opposite between the model result and the lattice data for $T/T_{\rm d}=1.68$ and $T/T_{\rm d}=2.10$. This discrepancy is also confirmed in Fig.~\ref{pxpz1} and the right panel of Fig.~\ref{HighTRatio}. However, since the purpose of the present study is to investigate possible roles of the Polyakov loops on $\mathbb{T}^2\times\mathbb{R}^2$, we do not care about the reproduction of the lattice data at the quantitative level. As discussed in Sec.~\ref{sec:level2}, as far as we have checked we could not find the parameter set that covers the whole $T$ range where the lattice data of Ref.~\cite{Kitazawa:2019otp} are available. This implies the necessity of further elaboration of the model building, which we leave for future study.

\begin{figure}
\includegraphics[scale=0.25]{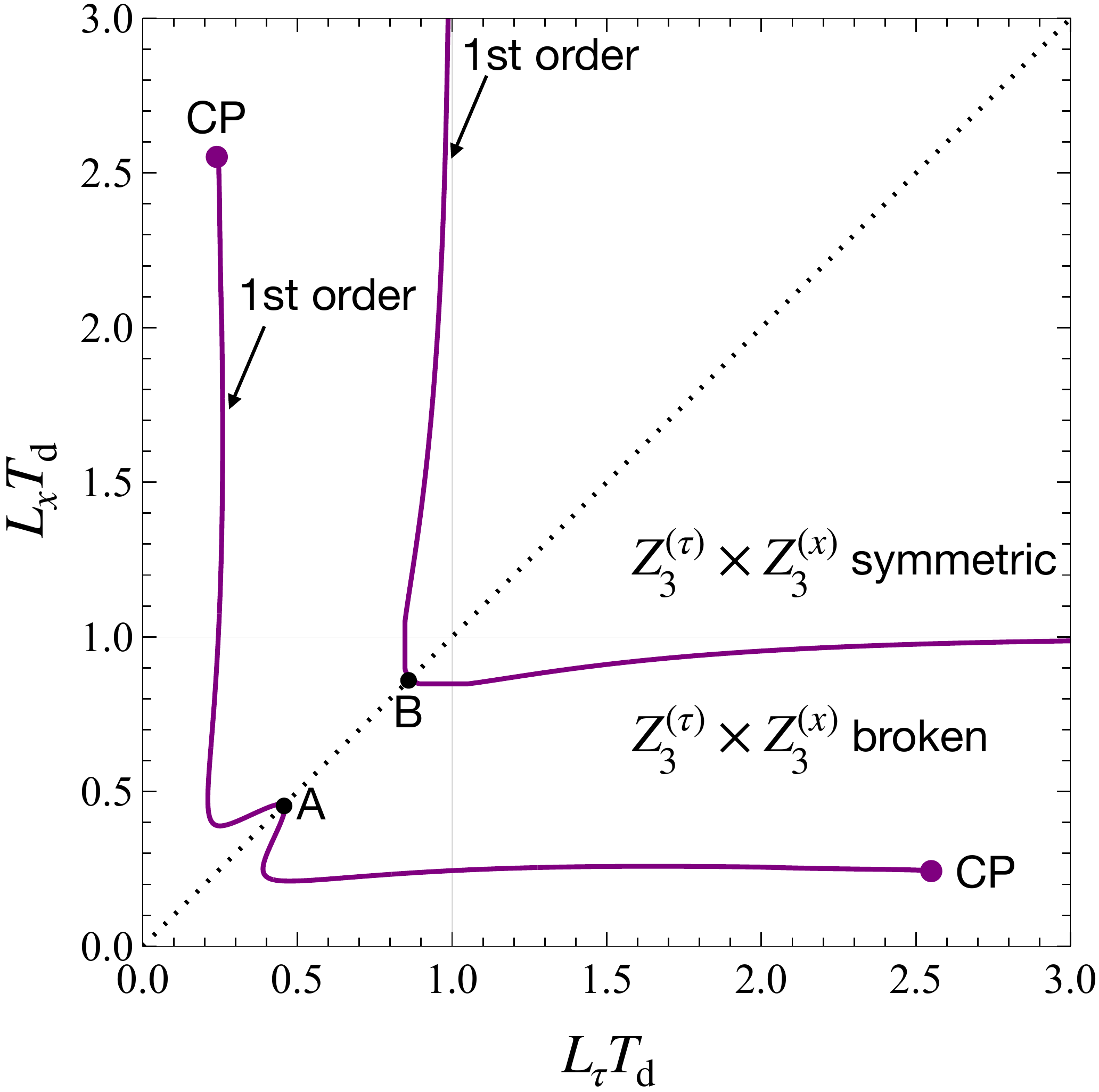}
\caption{
Phase diagram on the $L_\tau$--$L_x$ plane. The solid lines represent first-order phase transitions at which the thermodynamic quantities as well as $\Omega_c$ jump discontinuously. The critical points (CPs) are indicated by the big circles.}
\label{fig:PhaseDiagram} 
\end{figure}

\begin{figure}
\includegraphics[width=0.4\textwidth]{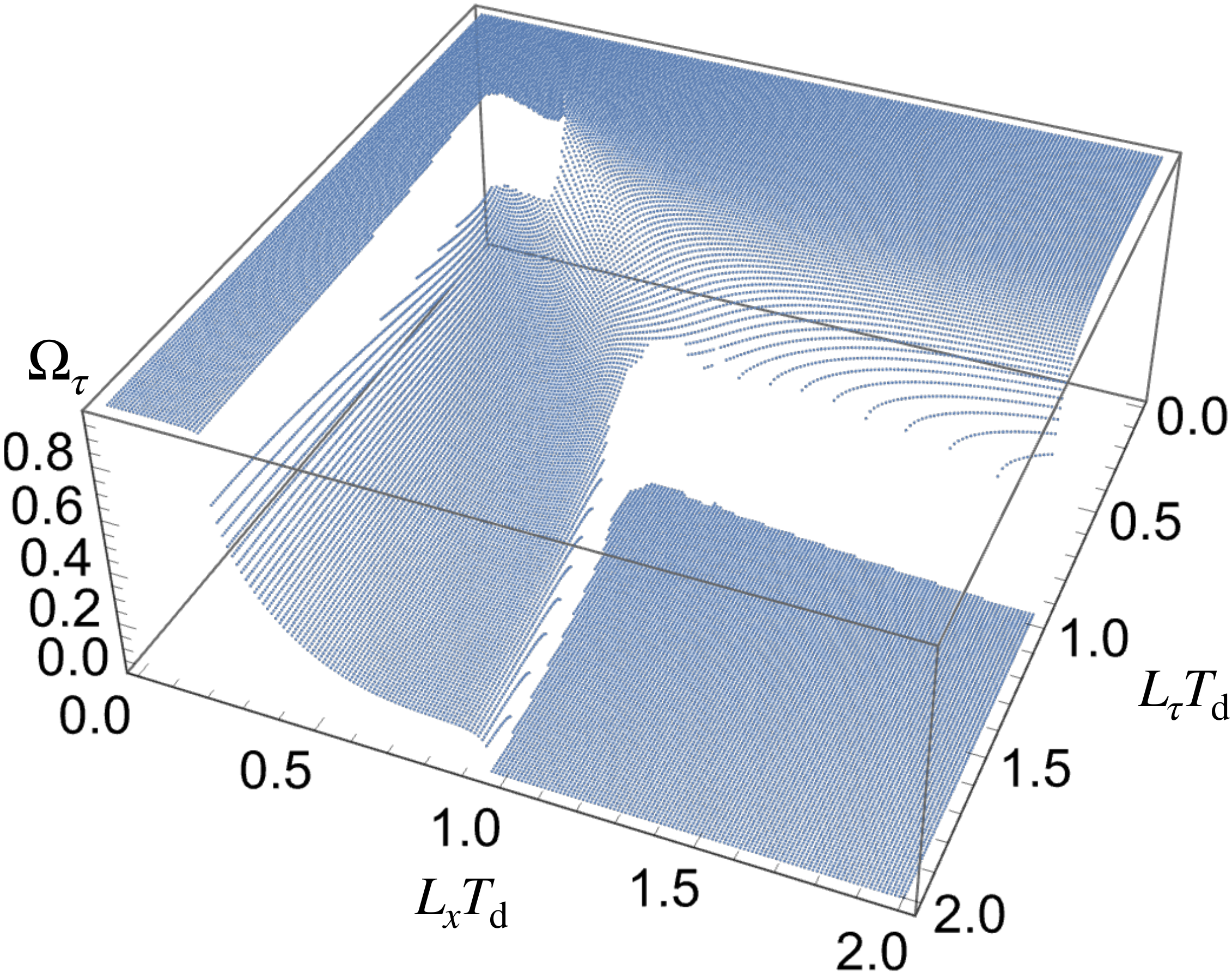} \\
\includegraphics[width=0.4\textwidth]{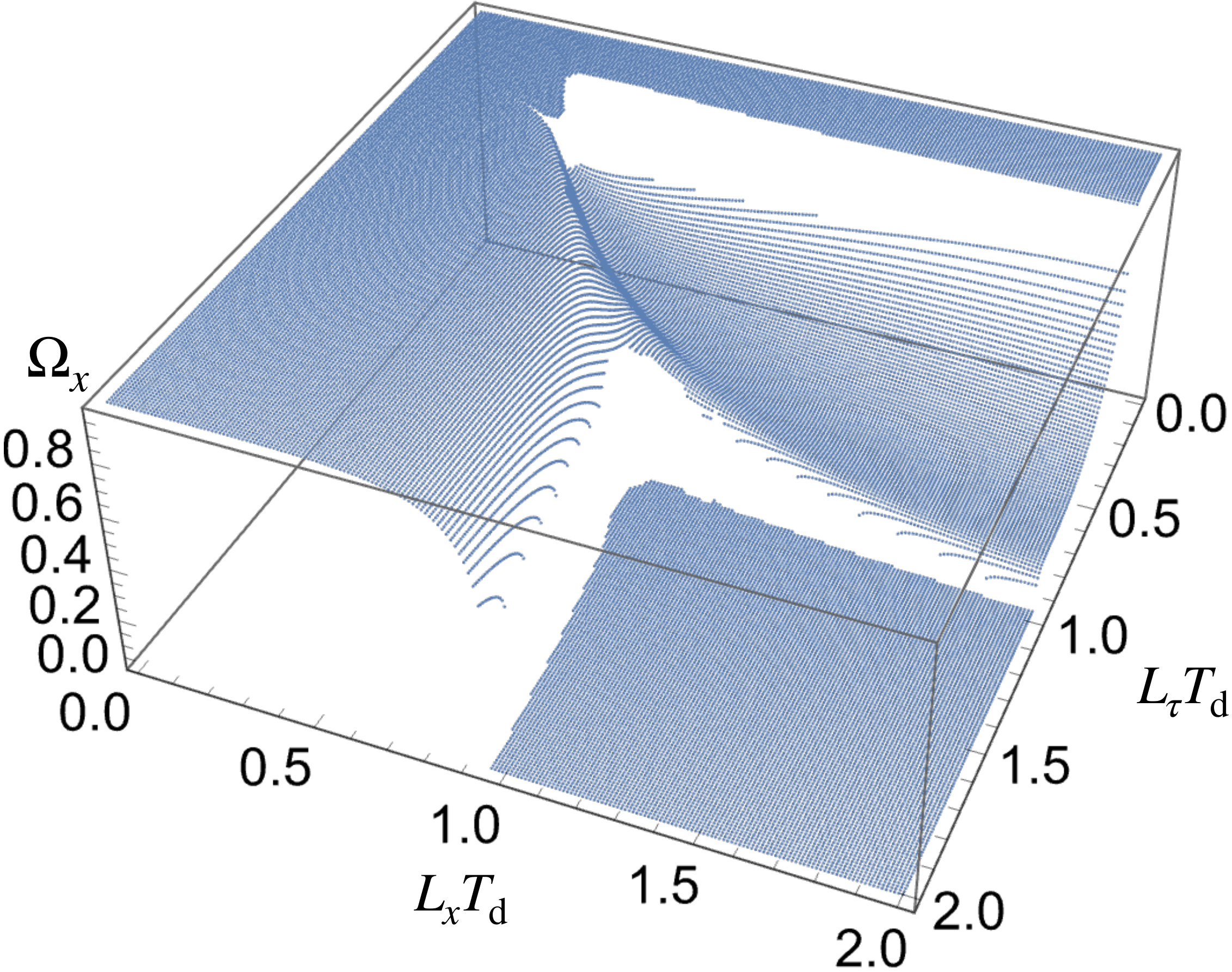}
\caption{Behaviors of $\Omega_\tau$ (top) and $\Omega_x$ (bottom) on the $L_\tau$--$L_x$ plane.}
\label{fig:PtauPx2DPlot.pdf} 
\end{figure}

\subsection{Phase diagram}
\label{sec:PhaseStructure}

In Fig.~\ref{fig:PhaseDiagram}, we depict the phase diagram of our model on the $L_\tau$--$L_x$ plane. The solid lines indicate the first-order phase transitions at which the thermodynamic quantities change discontinuously. The phase diagram is symmetric across $L_\tau=L_x$ indicated by the dotted line because of the transposition symmetry of $\tau$ and $x$ axes. In Fig.~\ref{fig:PtauPx2DPlot.pdf}, we also show the behaviors of $\Omega_\tau$ (top) and $\Omega_x$ (bottom) on the $L_\tau$--$L_x$ plane. 

One sees from these figures that there are two first-order transition lines on the $L_\tau$--$L_x$ plane, where $\Omega_\tau$ and $\Omega_x$ jump discontinuously in addition to the thermodynamic quantities~\footnote{Both $\Omega_\tau$ and $\Omega_x$ jump at the first-order transition line, although one of them is difficult to verify in Fig.~\ref{fig:PtauPx2DPlot.pdf}. See Fig.~\ref{fig:PtauPx_168_zoom.pdf} for an enlarged view of the jump.}. The line including the point B in Fig.~\ref{fig:PhaseDiagram} is connected to the confinement phase transition on $\SRRR$ in the large $L_x$ ($L_\tau$) limit at $L_\tau=1/T_{\rm d}$ ($L_x=1/T_{\rm d}$). As seen from Fig.~\ref{fig:PtauPx2DPlot.pdf}, the upper-right region of this line is the confined phase, where both $Z_3^{(c)}$ are restored with $\Omega_\tau=\Omega_x=0$. On the other hand, at the lower-left region both $Z_3^{(c)}$ are spontaneously broken with $\Omega_\tau\ne0$ and $\Omega_x\ne0$. Our model analysis shows that the phases where only one of the $Z_3^{(c)}$ is spontaneously broken do not appear on the phase diagram.

The other first-order transition line including the point A corresponds to the one found in Fig.~\ref{HighTRatio}.
As seen from Figs.~\ref{fig:PhaseDiagram} and~\ref{fig:PtauPx2DPlot.pdf}, this transition line lies entirely on the $Z_3^{(\tau)}\times Z_3^{(x)}$ broken phase. Moreover, the line terminates at finite $L_\tau$ and $L_x$ at 
$(L_\tau T_{\rm d},L_x T_{\rm d})\simeq(2.54,0.25)$ 
and $(0.25,2.54)$, and is not connected to any transitions on $\SRRR$.
The endpoint of a first-order transition line is the critical point (CP) at which the phase transition is of second order. The universality class of these CPs is specified as that of the two-dimensional Ising model ($Z_2$ universality class) as follows. First, on the first-order transition line two phases characterized by different $(\Omega_\tau,\Omega_x)$ coexist. Second, as the system approaches the CP the correlation length grows and eventually exceeds $L_x,\ L_\tau$. The system then can be regarded as two-dimensional. 

In QCD, CPs are known to manifest themselves with variations of various parameters such as the quark chemical potentials and the quark masses~\cite{Asakawa:1989bq,Kitazawa:2002jop,Philipsen:2021qji}. It is interesting that a novel existence of the CP is also indicated in $SU(3)$ YM theory, which is the heavy-mass limit of QCD, with the variations of $L_\tau$ and $L_x$.

\subsection{Phase transition on the $L_\tau=L_x$ line}
\label{sec:Ltau=Lx}

\begin{figure}
\includegraphics[width=0.45\textwidth]{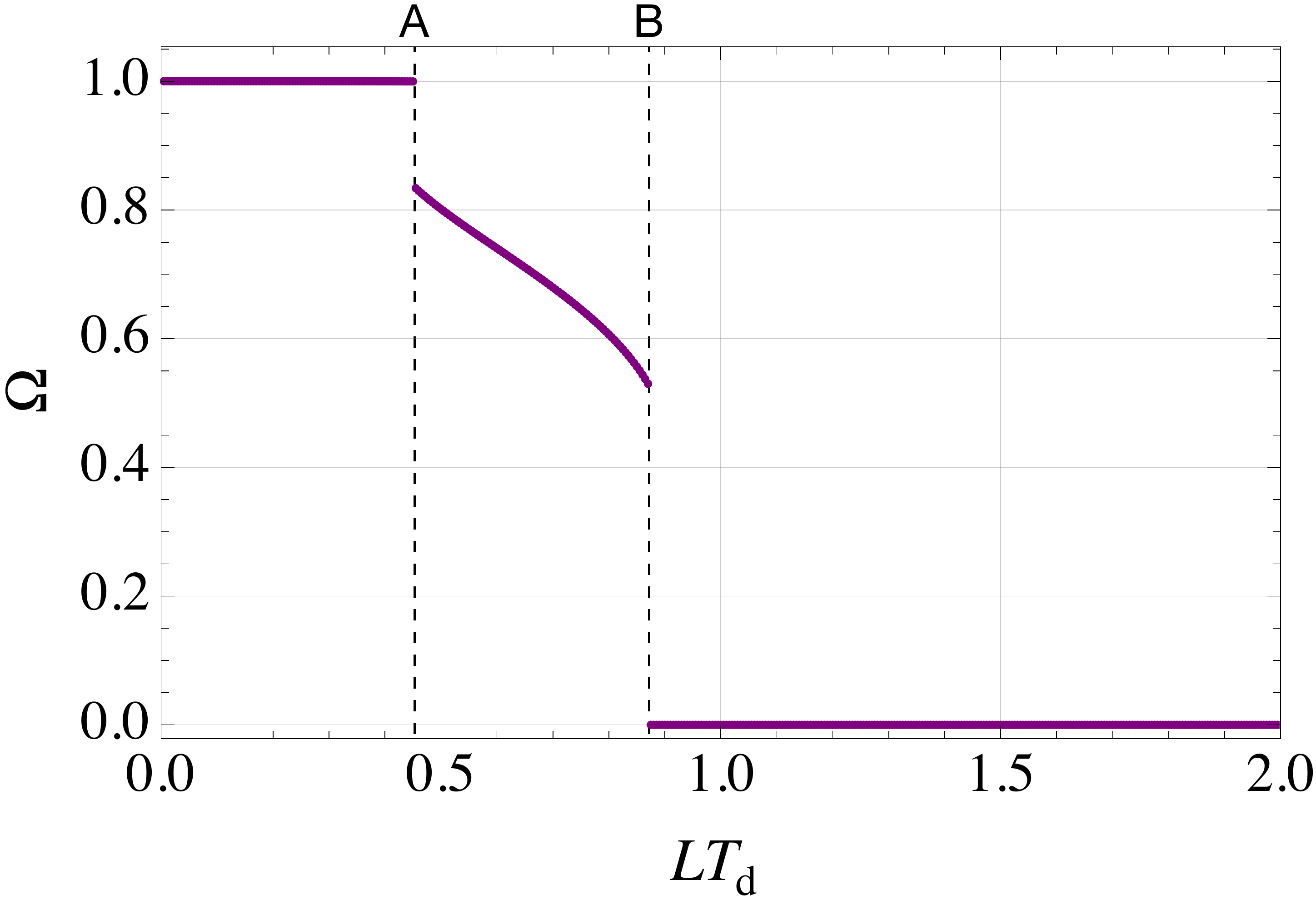}
\caption{Polyakov loop $\Omega=\Omega_\tau=\Omega_x$ on the symmetric trajectory $L=L_\tau=L_x$. }
\label{fig:PtauPx}
\end{figure}

In order to confirm 
the emergence of the novel first-order phase transition on $\TTRR$ more clearly, now let us investigate the phase transitions on the symmetric trajectory along $L_\tau=L_x$, i.e. the dotted line in Fig.~\ref{fig:PhaseDiagram}.

For $L_\tau=L_x$, the system is invariant under the transposition of $\tau$ and $x$ axes. As a result, $\Omega_\tau=\Omega_x$ (and hence $\phi_\tau=\phi_x$) is satisfied on this line as long as the transpose symmetry is not spontaneously broken. We have numerically verified that this is always the case in our model, although its violation is not prohibited in general. In Fig.~\ref{fig:PtauPx}, we show the behavior of $\Omega=\Omega_\tau=\Omega_x$ as a function of $L=L_\tau=L_x$. As in the figure, the value of $\Omega$ jumps at two points $L_{\rm A}T_{\rm d}\simeq0.455$ and $L_{\rm B}T_{\rm d}\simeq0.87$, corresponding to the points A and B in Fig.~\ref{fig:PhaseDiagram}.

On the $L_\tau=L_x$ line, the free-energy density~\eqref{eq:f} is written as
\begin{align}
\tilde f(\phi;L) 
= &\, f(\vec\theta,\vec\theta;L,L)\big|_{\vec\theta=(\phi,0,-\phi)}
\notag \\
= &\, 
\tilde f_{\rm pert+sep}(\phi;L) 
+ \tilde f_{\rm cross}^{c_4}(\phi;L)  
\notag \\
& + \tilde f_{\rm cross}^{c_5}(\phi;L) + \tilde f_{\rm cross}^{c_6}(\phi;L) ,
\label{eq:ftilde}
\end{align}
with $\phi=\phi_\tau=\phi_x$, where $\tilde f_{\rm pert+sep}$ is the contribution from Eqs.~\eqref{fpert1} and~\eqref{eq:fsep}, and $\tilde f_{\rm cross}^{c_4}$, $\tilde f_{\rm cross}^{c_5}$, $\tilde f_{\rm cross}^{c_6}$ represent the terms in Eq.~\eqref{eq:fcross} containing $c_4,\ c_5,\ c_6$, respectively.
From Eqs.~\eqref{eq:Omega1phi} and~\eqref{eq:Omega3phi}, one finds
\begin{align}
\tilde f_{\rm cross}^{c_4}(\phi;L) =\,& c_4 g(L,L) (1+2\cos\phi)^4 ,
\label{eq:f4}
\\
\tilde f_{\rm cross}^{c_5}(\phi;L) =\,& c_5 g(L,L) (1+2\cos\phi)^2 (1+2\cos(3\phi)),
\label{eq:f5}
\\
\tilde f_{\rm cross}^{c_6}(\phi;L) =\,& c_6 g(L,L) (1+2\cos(3\phi))^2.
\label{eq:f6}
\end{align}
These formulas show that $\tilde f_{\rm cross}^{c_4}(\phi;L)$ and $\tilde f_{\rm cross}^{c_5}(\phi;L)$ have a minimum at $\phi=2\pi/3$ and $\phi\simeq\pi/3$, respectively, provided $c_i>0$, while $\tilde f_{\rm cross}^{c_6}(\phi;L)$ has five extrema at $\phi=0,\ 2\pi/9,\ \pi/3,\ 4\pi/9,\ 2\pi/3$.

\begin{figure*}
\includegraphics[scale=0.3]{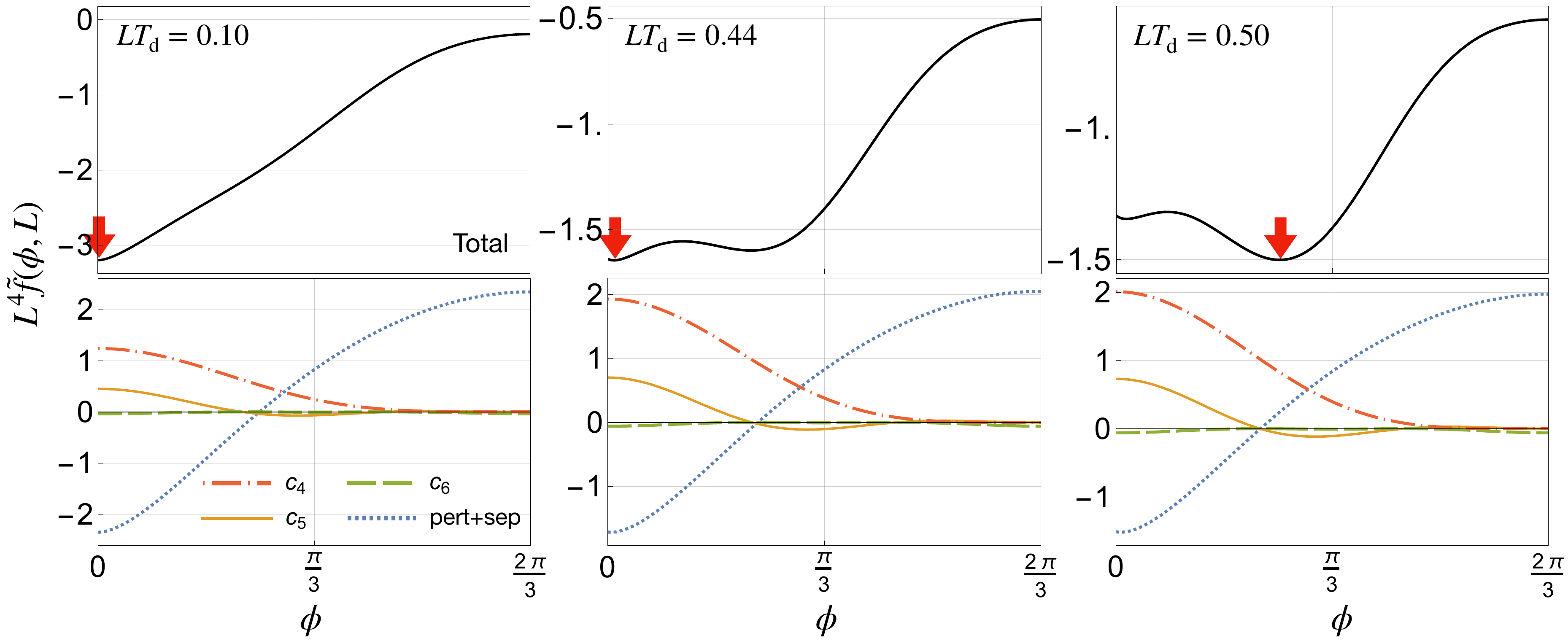}
\caption{Total free energy $\tilde{f}$ (upper panels) and separate contributions in Eq.~\eqref{eq:ftilde} (lower panels) as functions of $\phi$ at several values of $LT_{\rm d}$. $L^4$ is used to make the vertical axis dimensionless. $\tilde{f}_{\rm pert+sep}$ is plotted with a constant shift for presentation.}
\label{fig:free_energy2} 
\end{figure*}

In the upper panels of Fig.~\ref{fig:free_energy2}, we show the free-energy density $\tilde f(\phi;L)$ at $LT_{\rm d}=0.1$, $~0.44,~0.50$.
In the lower panels, the components in Eq.~\eqref{eq:ftilde} are plotted separately. 
The red arrows represent the global minimum of $\tilde f(\phi;L)$. 

The top-left panel shows that $\phi\simeq0$ ($\Omega\simeq1$) is favored as the global minimum at $LT_{\rm d}=0.10$. In the top-middle and top-right panels, one sees that another local minimum emerges as $L$ becomes larger and it eventually becomes the global one at $LT=0.5$. This leads to the discontinuous jump of $\phi$ at $L=L_{\rm A}$. Although the other first-order transition occurs at $L=L_{\rm B}$, we do not discuss it here in detail since the point is outside the applicable range of our model, as discussed already.

The lower panels of Fig.~\ref{fig:free_energy2} allow us to understand the roles of $f_{\rm cross}^{c_i}$. First, these panels show that $f_{\rm cross}^{c_4}$ acts to favor the confined phase with large $\phi$ (small $\Omega$). This feature is directly seen in Eq.~\eqref{eq:f4}. As we will see in the next section, this term plays a crucial role in realizing the flat behaviors of $p_x/p_z$ at $L_xT\gtrsim1.3$ in Fig.~\ref{pxpz1}. Next, since $f_{\rm cross}^{c_5}$ has a minimum at $\phi\simeq\pi/3$, this term with $c_5>0$ acts to trap the value of $\phi$ around there. 
As discussed in App.~\ref{sec:parameterdep}, the location of the point A is sensitive to $c_5$. 
This term is also indispensable in reproducing the lattice data at $L_xT\lesssim 1.3$ in Figs.~\ref{pxpz1} and~\ref{epxpz}. 
Finally, the role of $c_6$ is smaller than the other two terms, while this term is also important to determine the thermodynamics near $L_\tau=L_x$.

\begin{figure}
    \centering
    \includegraphics[width=0.46\textwidth]{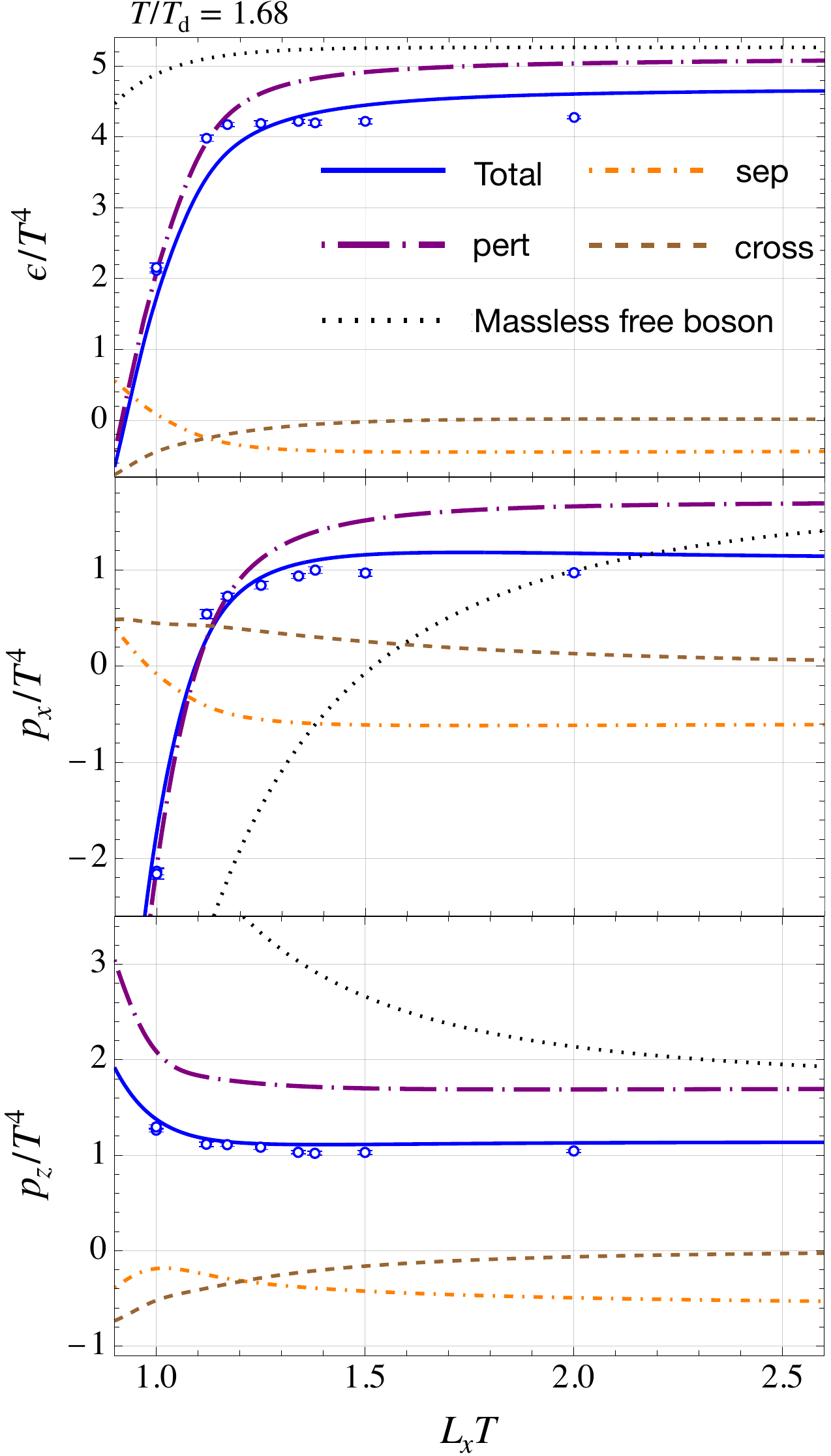}
    \caption{Decompositions of $\epsilon$, $p_x$, and $p_z$ into three parts as Eqs.~\eqref{eq:edecomp}--\eqref{eq:pzdecomp} at $T/T_{\rm d}=1.68$. The circle markers shows the lattice data. }
    \label{fig:decomp}
\end{figure}

\section{Effect of Polyakov loops on thermodynamics}
\label{sec:physical}

In the previous section, we have seen that the flat behavior $p_x/p_z\simeq1$ for $L_xT\gtrsim1.3$ observed on the lattice is well reproduced in our model. In this section, we investigate the mechanism that favors this behavior in our model in more detail.

To begin with, we note that the thermodynamic quantities in our model are decomposed as
\begin{align}
    \epsilon =&\, \epsilon^{\rm pert} + \epsilon^{\rm sep} + \epsilon^{\rm cross} ,
    \label{eq:edecomp}
    \\
    p_x =&\, p_x^{\rm pert} + p_x^{\rm sep} + p_x^{\rm cross} ,
    \label{eq:pxdecomp}
    \\
    p_z =&\, p_z^{\rm pert} + p_z^{\rm sep} + p_z^{\rm cross} ,
    \label{eq:pzdecomp}
\end{align}
where the three terms on the right-hand side correspond to $f_{\rm pert}$, $f_{\rm sep}$ and $f_{\rm cross}$ through Eqs.~\eqref{Energy}--\eqref{PressureZ}. In Eqs.~\eqref{eq:edecomp} and~\eqref{eq:pxdecomp}, we define the $L_c$ derivatives with fixed $\Omega_\tau$ and $\Omega_x$ as 
\begin{align}
    \epsilon^{\rm pert} 
    = \frac{\partial (L_\tau f_{\rm pert})}{\partial L_\tau}\Big|_{\Omega_\tau,\Omega_x} ,
    \quad
    p_x^{\rm pert} 
    = -\frac{\partial (L_x f_{\rm pert})}{\partial L_x}\Big|_{\Omega_\tau,\Omega_x} ,
    \label{eq:epert}
\end{align}
and etc., i.e. we do not take account of derivatives acting on $\Omega_c$, although the decomposition including them is also possible. Since $\partial f/\partial\Omega_c=0$ is satisfied for the total free energy $f$ from the stationary conditions, the decompositions~\eqref{eq:edecomp}--\eqref{eq:pzdecomp} are valid for both the definitions.

In Fig.~\ref{fig:decomp}, we show individual components in Eqs.~\eqref{eq:edecomp}--\eqref{eq:pzdecomp} for $T/T_{\rm d}=1.68$. The dotted line in each panel is the result of the massless-free boson system. From these results, one sees that the perturbative contributions $\epsilon^{\rm pert}$, $p_x^{\rm pert}$, and $p_z^{\rm pert}$ solely have the same trend as the lattice data. This suggests that these terms are responsible for the reproduction of the lattice data. A similar result is obtained for $T/T_{\rm d}=2.10$.

\begin{figure}
\includegraphics[scale=0.18]{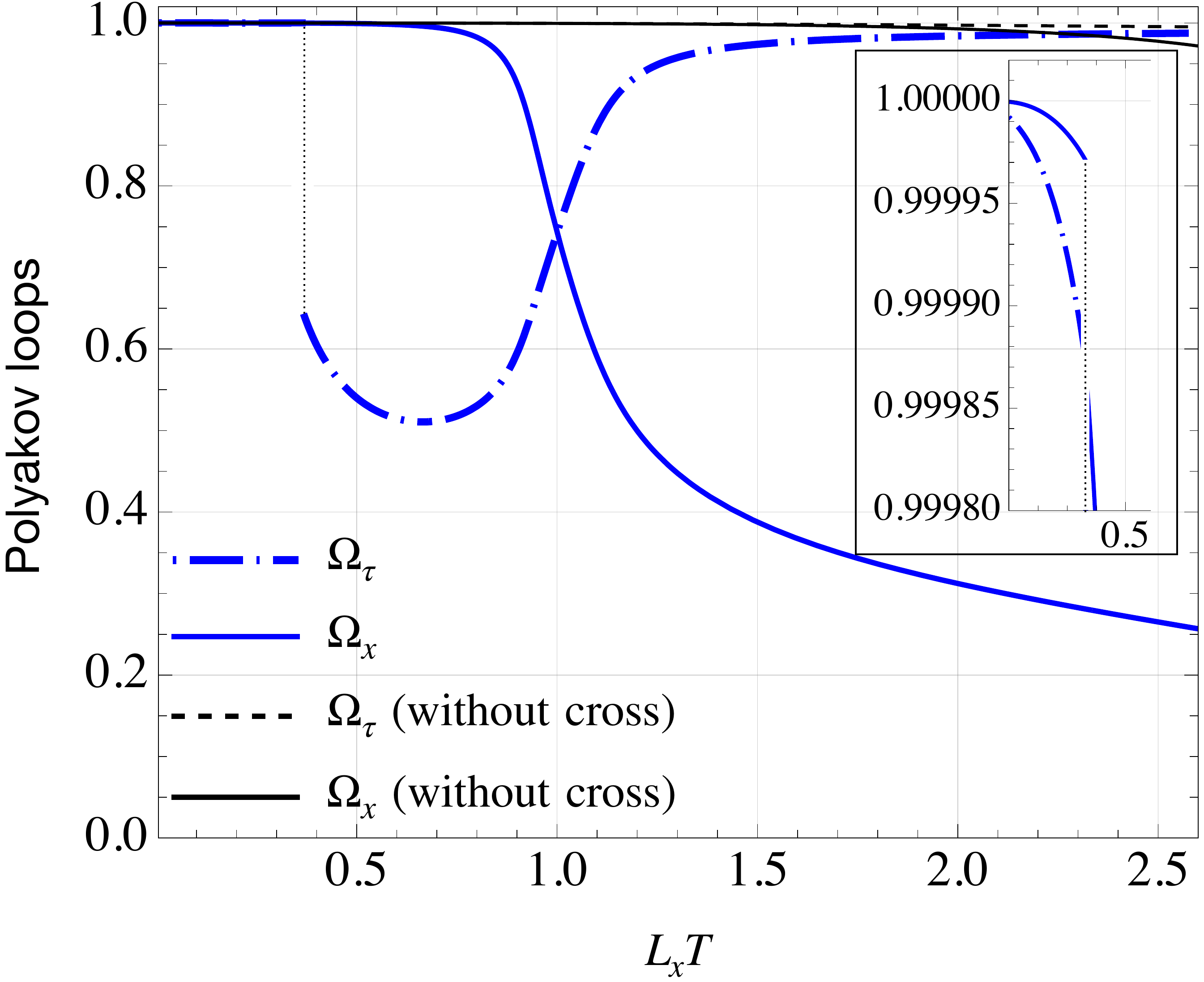}
\caption{Dependence of the Polyakov loop $\Omega_\tau$ and $\Omega_x$ on $L_xT$ at $T/T_{\rm d}=1.68$. The thin lines show the result obtained without the cross term $f_{\rm cross}$.}
\label{fig:PtauPx_168_zoom.pdf} 
\end{figure}

Next, in Fig.~\ref{fig:PtauPx_168_zoom.pdf} we show the behaviors of $\Omega_\tau$ and $\Omega_x$ as functions of $L_xT$ at $T/T_{\rm d}=1.68$. From the figure, one sees that $\Omega_x$ takes a small value at $L_xT\gtrsim1.3$, and suddenly approaches unity at $0.8\lesssim L_xT\lesssim1.3$. In the figure, we also show the $L_xT$ dependence of $\Omega_\tau$ and $\Omega_x$ when $f_{\rm cross}$ is switched off by the thin-black lines for comparison, which approximately reproduces Ref.~\cite{Suenaga:2022rjk}. One sees that in this case $\Omega_\tau\sim1$ and $\Omega_x\sim1$ are satisfied at the whole range of $L_xT$ shown in the figure. 
From this difference, it is deduced that the suppression of $\Omega_x$ is responsible for the reproduction of the lattice data.
From the lower panels of Fig.~\ref{fig:free_energy2}, it is also understood that the suppression of $\Omega_x$ due to $f_{\rm cross}$ is predominantly attributed to the term proportional to $c_4$. 

Effects of $\Omega_x$ on $p_x^{\rm pert}$ and $p_z^{\rm pert}$ are analytically understood as follows. 
From Eq.~\eqref{fpert1}, $p_x^{\rm pert}$ is calculated to be
\begin{align}
    & p_x^{\rm pert} 
    =  - \frac{\partial(L_x f_{\rm pert})}{\partial L_x} \Big|_{\Omega_\tau,\Omega_x}
    \notag \\
    &= - \frac{1}{L_\tau}\sum^3_{j,k=1}\Big(1-\frac{\delta_{jk}}{3}\Big)\sum_{\ell_\tau}\int \frac{d^2p_L}{(2\pi)^2} n_{L_x}({\cal E},(\Delta\theta_x)_{jk}),
    \label{eq:pxpert}
\end{align}
with
\begin{align}
    n_\beta(E,\varphi) =&\  E \Big( 1 + \frac1{e^{\beta E+i\varphi}-1}+ \frac1{e^{\beta E-i\varphi}-1} \Big),
    \label{eq:n}
    \\
    {\cal E}^2 = & \ (\omega_\tau-(\Delta\theta_\tau)_{jk}/L_\tau)^2+\bm{p}^2_L ,
\end{align}
where we used the identity
\begin{align}
    \frac\partial{\partial\beta} \sum_{l\in \mathbb{Z}} \ln \Big[ ( \omega_l - \varphi/\beta)^2 + E^2 \Big] = n_\beta(E,\varphi),
    \label{eq:lnsum}
\end{align}
with the Matsubara modes $\omega_l=2\pi l/\beta$~\footnote{The first term in Eq.~\eqref{eq:n} is customarily subtracted in the analysis of thermodynamics so that they vanish in the vacuum.}. 
From Eq.~\eqref{eq:pxpert} and the fact that $n(E,\varphi)$ is a monotonically-decreasing function of $|\varphi|$ at $|\varphi|\le\pi$, one finds that $p_x^{\rm pert}$ is an increasing function of $\phi_x$ due to the overall minus sign. In other words, smaller $\Omega_x$ enhances $p_x$. 
Moreover, thermodynamics is insensitive to $\Omega_x$ for $L_xT\gg1$, where the effects of PBC along $x$ direction is negligible. This explains the reason why $p_x$ is significantly modified in the range of $L_xT$ shown in Fig.~\ref{fig:decomp}. 

A similar argument is also applicable to $p_z$.
By integrating Eq.~\eqref{eq:pxpert} with respect to $L_x$ after removing the constant term in Eq.~\eqref{eq:n}, one obtains 
\begin{align}
    p_z^{\rm pert} 
    = & - \frac\partial{\partial L_z} (L_z f_{\rm pert})
    \notag \\
    = & - \ \frac{1}{L_\tau}\sum^3_{j,k=1}\Big(1-\frac{\delta_{jk}}{3}\Big)\sum_{\ell_\tau}\int \frac{d^2p_L}{(2\pi)^2} 
    \notag \\
    & \times \ln( 1 - e^{-\beta {\cal E}+i(\Delta\theta_x)_{jk}}) ( 1 - e^{-\beta {\cal E}-i(\Delta\theta_x)_{jk}}) .
    \label{eq:pzpert}
\end{align}
From the fact that the logarithmic term in Eq.~\eqref{eq:pzpert} is a decreasing function of $(\Delta\theta_x)_{jk}$, one can conclude that smaller $\Omega_x$ suppresses $p_z$. 

Since the suppression of $\Omega_x$ leads to an enhancement of $p_x$ and suppression of $p_z$, it leads to larger $p_x/p_z$ than the free-boson system, which explains the result in Fig.~\ref{pxpz1}.

\section{Conclusion and outlook}

In this paper, we have investigated the thermodynamics of $SU(3)$ Yang-Mills theory on $\mathbb{T}^2\times\mathbb{R}^2$ using an effective model including two Polyakov loops along two compactified directions, $\Omega_\tau$ and $\Omega_x$, as dynamical variables.
We extended the model employed in Ref.~\cite{Suenaga:2022rjk} by introducing the cross terms in the Polyakov-loop potential, which physically represent their interplay. We found that our model can successfully reproduce the qualitative behavior of thermodynamics measured on the lattice~\cite{Kitazawa:2019otp} for the temperature range $T/T_{\rm d}\gtrsim1.5$, while the lattice data for $T/T_{\rm d}\lesssim1.5$ and small $L_xT$ are difficult to reproduce.

An interesting outcome of this study is that our model predicts the manifestation of a novel first-order phase transition on $\TTRR$. The transition appears in the $Z_3^{(\tau)}\times Z_3^{(x)}$ broken phase with $\Omega_\tau\ne0$ and $\Omega_x\ne0$, and is not connected to any phase transitions on $\SRRR$. Moreover, the existence of the critical points as the endpoints of the first-order transition line, which should belong to the two-dimensional $Z_2$ universality class, is suggested. We have also elucidated the mechanism for the emergence of the first-order transition and the flat $p_x/p_z$ behavior observed in Ref.~\cite{Kitazawa:2019otp}. 

The existence of the novel phase transitions on $\TTRR$ predicted in the present study can be verified straightforwardly in lattice numerical simulations. While the data in Ref.~\cite{Kitazawa:2019otp} have sharp variations as in Fig.~\ref{HighTRatio}, the data points are still coarse to give a definite conclusion. Our results strongly motivate the improvement of these data. Also, the simultaneous measurement of the thermodynamic quantities and the Polyakov loops in these analyses is highly desirable to understand their mutual roles.
Improving the lattice data at low-temperature part $T\simeq T_{\rm d}$ will also be useful, whereas the model has a poor agreement with the lattice data there. 

There are many possible extensions of the present study. Although we focused on $SU(3)$ YM theory motivated by the available lattice data on $\TTRR$, similar analysis in other theories is also interesting. In particular, $SU(2)$ YM theory is promising since the analysis of the lattice suggests an interesting phase structure on $\TTRR$~\cite{Chernodub:2017mhi,Chernodub:2018aix}. The large-$N$ gauge theories and theories including fermions are other interesting applications. In particular, in the application of the present study to relativistic heavy-ion collisions, clarifying the role of fermions would be crucial. To understand the phase structure on $\TTRR$, it is also interesting to explore the YM theory in $2+1$ dimensions, since it is the $L_x\to0$ limit in the phase diagram on the $L_\tau$--$L_x$ plane.
Whereas we introduced the potential term in a phenomenological manner in the present study, it is interesting to pursue its derivation based on a theoretical treatment, such as the perturbation theory and AdS/CFT correspondence. We leave these subjects for future studies.

\section*{Acknowledgments}

The authors thank Takeshi Morita, Akira Ohnishi, Robert D. Pisarski, and Kei Suzuki for useful discussions.
D.S. was supported by the RIKEN special postdoctoral researcher program. This work was supported in part by the Japan Society for the Promotion of Science (JSPS) KAKENHI Grants Nos.~19H05598, 22K03619, 23K03377, 23H04507, 23H05439 and by the Center for Gravitational Physics and Quantum Information (CGPQI) at Yukawa Institute for Theoretical Physics.

\appendix

\section{Assumptions in the model}
\label{ansatz}

In this Appendix, we summarize the assumptions and ans\"atze imposed in the model construction in this study. Since the model is constructed as an extension of Refs.~\cite{Meisinger:2001cq,Dumitru:2012fw}, we discuss them starting from these studies.

In our model, in response to the compactification of the $x$ direction we introduce the spatial Polyakov loop $\Omega_x(\bm{x}_x^\perp)$ as a dynamical variable in addition to the temporal one $\Omega_\tau(\bm{x}_\tau^\perp)$ into the model in Refs.~\cite{Meisinger:2001cq,Dumitru:2012fw}. Here, $\Omega_x(\bm{x}_x^\perp)$ is treated as a constant field independent of $\bm{x}_x^\perp$ as $\Omega_\tau(\bm{x}_\tau^\perp)$ in Refs.~\cite{Meisinger:2001cq,Dumitru:2012fw}. These assumptions are the basis of our effective model. The successful reproduction of the lattice thermodynamics in Refs.~\cite{Meisinger:2001cq,Dumitru:2012fw} may support its validity on $\SRRR$. Our model also inherits some other assumptions from Refs.~\cite{Meisinger:2001cq,Dumitru:2012fw}. For example, Eq.~\eqref{theta} is not the most general form of the gauge field. Their validity may be verified from the comparison of the model results with the lattice data. On $\TTRR$, it is also assumed that the two Polyakov loops $\Omega_\tau(\bm{x}_\tau^\perp)$ and $\Omega_x(\bm{x}_x^\perp)$ are diagonalized simultaneously. This assumption is necessary to represent the free energy~\eqref{fpert2} in a compact form. 

To define $f_{\rm pot}(\vec\theta_\tau,\vec\theta_x;L_\tau,L_x)$ as a function of two Polyakov loops, we assume that this function is given by the form~\eqref{FPotSeparate}. This functional form is introduced as a choice to satisfy the constraints~(i)--(iv) in Sec~\ref{sec:citeref}, but obviously it is not a unique choice. Whereas the constraints~(i), (iii), and~(iv) may be robust in general, the constraint~(ii) is an ans\"atz in our model construction and may be replaced with a better one in the future study. 

Among the constraints (v)--(viii) enumerated to determine the form of $f_{\rm cross}$, (v)--(vii) are direct consequences of (i)--(iv). On the other hand, in the constraint~(viii) we introduce a nontrivial assumption that $f_{\rm cross}$ is composed of terms containing $\Omega_c$ up to the third order as in Eq.~\eqref{eq:fcross}, although higher-order terms are not excluded from general arguments. Also, we assume \eqref{fTypeI} as a $L_\tau,L_x$-dependence of this term. However, this choice is a na\"ive assumption that is introduced to reduce the number of free parameters.

By altering these assumptions, our model would be improved further. Since our model fails in reproducing the lattice data for $T\lesssim1.5T_{\rm d}$, it is especially interesting to improve the low-$T$ behavior by modifying the assumptions.

\section{Double summation}
\label{doublesum}

In this Appendix, we derive Eq.~\eqref{doublesumconcrete}. 
We start from
\begin{align}
    \sum_{m=1}^\infty \frac{\cos(am)}{(m^2+C^2)^2}
=
\sum_{m=-\infty}^\infty \frac{e^{iam}}{2(m^2+C^2)^2}-\frac{1}{2C^4},
\label{eq:sum1}
\end{align}
with $C\ne0$.
The sum over $m$ on the right-hand side can be taken through 
a formula
\begin{align}
    \sum_{n=-\infty}^\infty R(n) e^{in\xi}
=
-\sum_{\{\zeta\}} \mathrm{Res}
\left(
R(z)\frac{2\pi ie^{iz\xi}}{e^{2\pi i z}-1};\zeta
\right) ,
\label{eq:sumR}
\end{align}
where $R(z)$ is any rational function which 
has no poles at $z=n\in\mathbb{Z}$ and 
drops faster than $|z|^{-1}$ for $|z|\to\infty$, $\{\zeta\}$ is the set of all poles of $R(z)$, and $\xi\in[0,2\pi)$. 
Substituting $R(n)=1/[2(n^2+C^2)^2]$
into Eq.~\eqref{eq:sumR} yields
\begin{align}
&\sum_{m=1}^\infty \frac{\cos(am)}{(m^2+C^2)^2}
\notag
\\=&
\frac{\pi}{4C^2}\frac{1}{\sinh(\pi C)}
\left[
\frac{\cosh[(\pi-a)C]}{C}
\right.\notag
\\&\left.
+a \sinh[(\pi-a)C]
+\pi\frac{\cosh(a C)}{\sinh(\pi C)}
\right]
-\frac{1}{2C^4}.
\label{eq:Sum1st}
\end{align}
Equation~\eqref{doublesumconcrete} is obtained straightforwardly using Eqs.~\eqref{eq:Sum1st} with $C=l_x(L_\tau/L_x)$.

\section{Role of each parameter} 
\label{sec:parameterdep}

\begin{figure}
\includegraphics[width=0.48\textwidth]{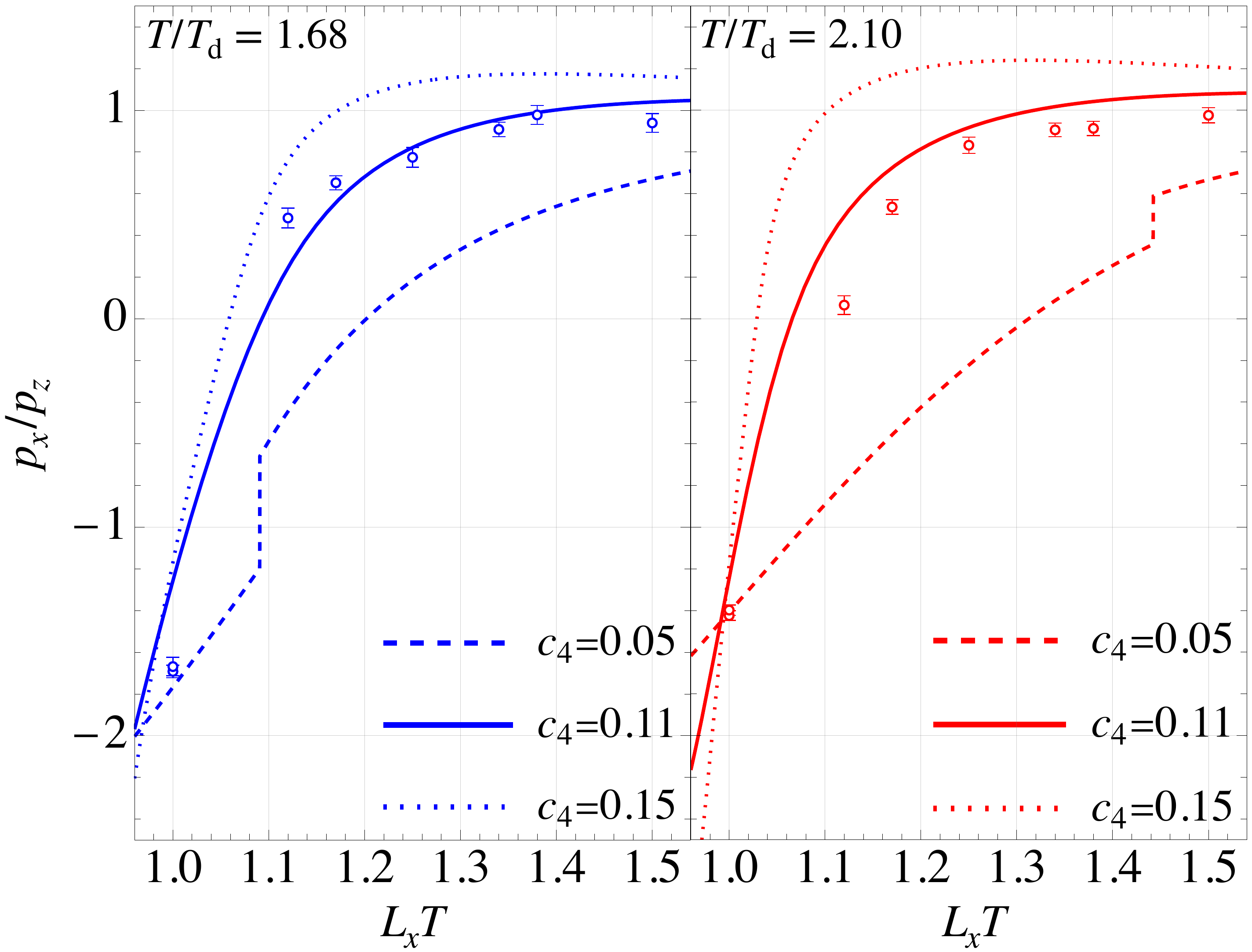}
\caption{\label{pxpzc4} $L_xT$ dependence of $p_x/p_z$ at $T/T_{\rm d}=1.68,\ 2.10$ for $c_4=0.05,\ 0.11,\ 0.15$. The other parameters are fixed to Eq.~\eqref{FCrossParameter}. The circle markers show the lattice data.}
\end{figure}

\begin{figure}
\includegraphics[width=0.48\textwidth]{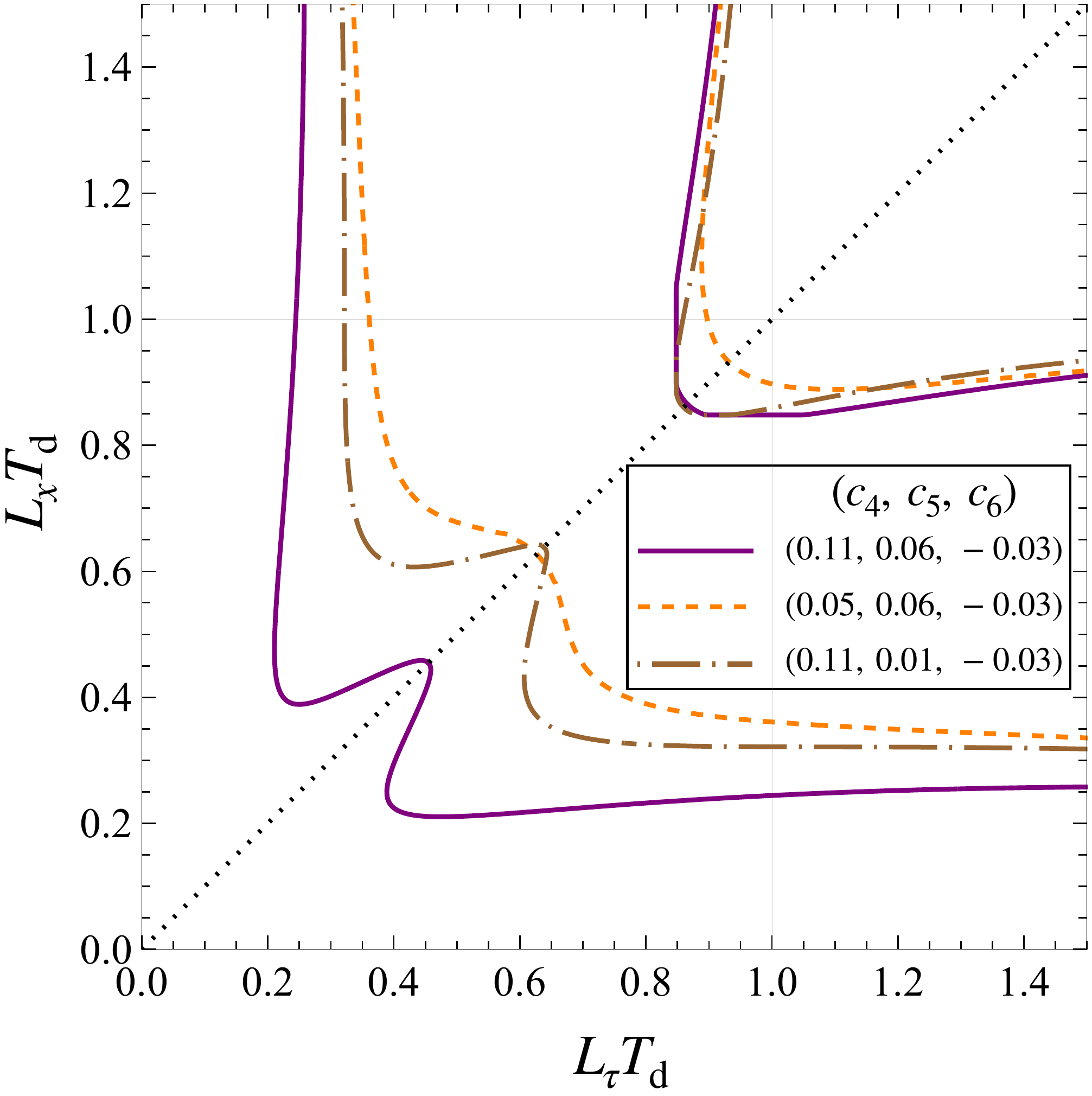}
\caption{\label{phasediagram_smallc45} Phase diagram on the $L_\tau$--$L_x$ plane for several parameter sets with $n=1.85$.} 
\end{figure}

\begin{figure}
\includegraphics[scale=0.18]{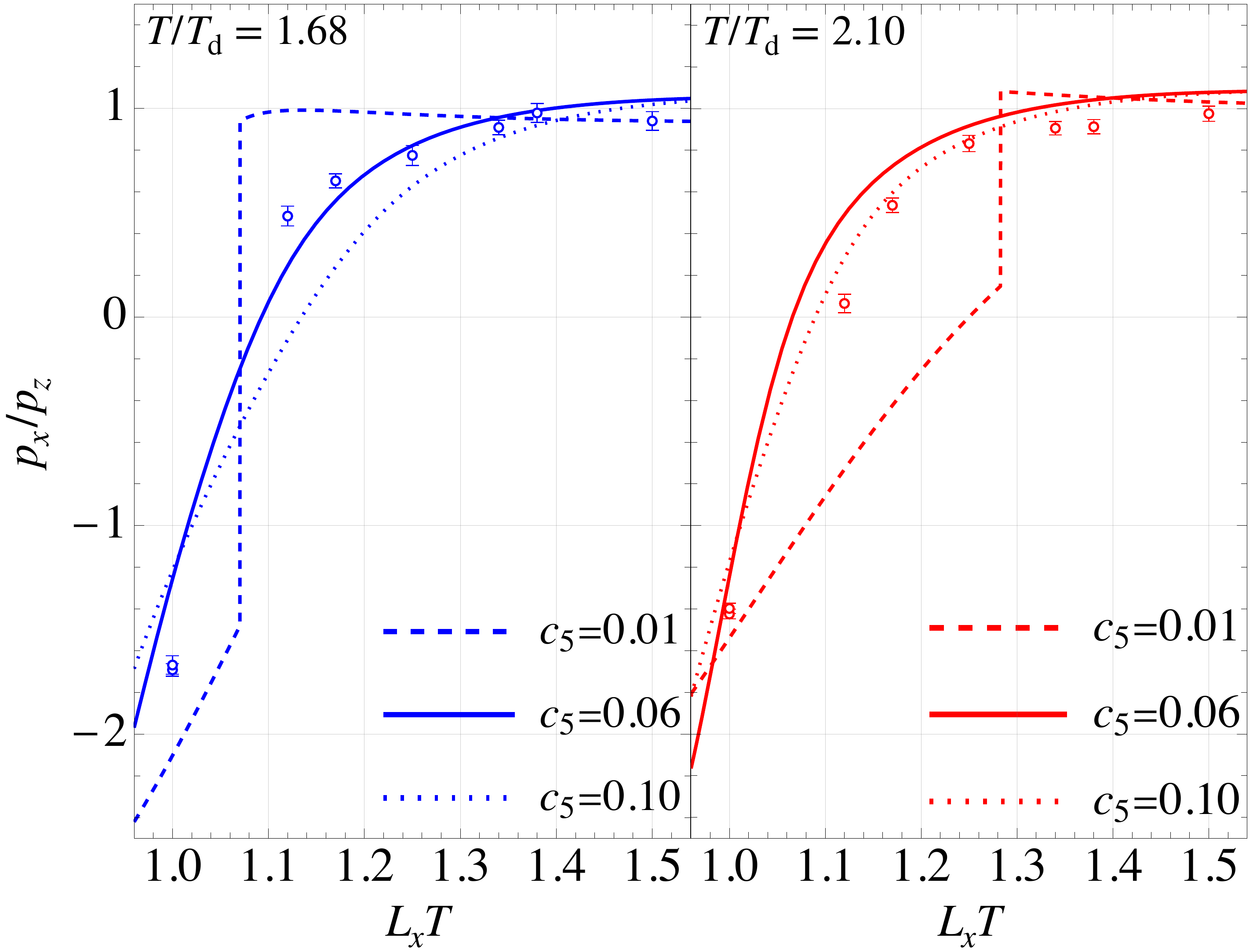}
\caption{\label{pxpzc5} $L_xT$ dependence of $p_x/p_z$ at $T/T_{\rm d}=1.68,\ 2.10$ for $c_5=0.01,\ 0.06,\ 0.10$. }
\end{figure}

In this Appendix, we discuss the dependence of thermodynamic quantities on the parameters in our model, $c_4,\ c_5,\ c_6,\ n$. This analysis allows us to understand the role of each parameter in our model, and clarifies the procedure to obtain the parameter set in Eq.~\eqref{FCrossParameter}.

In Fig.~\ref{pxpzc4}, we first show the $L_xT$ dependence of $p_x/p_z$ at $T/T_{\rm d}=1.68,\ 2.10$ for several values of $c_4$. The solid lines show the results with the parameter set in Eq.~\eqref{FCrossParameter}, while the dotted and dashed lines correspond to the results with $c_4=0.05$ and $0.15$, respectively, with the other parameters fixed. The figure shows that $p_x/p_z$ increases with decreasing $c_4$. This behavior is understood from the discussions in Secs.~\ref{sec:Ltau=Lx} and~\ref{sec:physical} as follows. As we have seen in Sec.~\ref{sec:Ltau=Lx}, $c_4$ acts to reduce the value of the Polyakov loops $\Omega_c$. Then, from the argument in Sec.~\ref{sec:physical} one finds that smaller $\Omega_x$ enhance $p_x/p_z$~\footnote{Precisely speaking, the variation of $c_4$ also affects thermodynamics through the last terms in Eqs.~\eqref{eq:edecomp}--\eqref{eq:pzdecomp}. As in Fig.~\ref{fig:decomp}, this contribution is not negligible.}. 

In Fig.~\ref{phasediagram_smallc45}, we depict the phase diagram on the $L_\tau$--$L_x$ plane for several parameter sets. The solid lines show the first-order transition for the parameter set~\eqref{FCrossParameter}, and the dashed lines represent its location for $c_4=0.05$ with the other parameters fixed. One finds that the first-order transition shifts toward the right-upper region by decreasing $c_4$. This behavior is nicely understood from the discussion in Sec.~\ref{sec:Ltau=Lx}.

In Fig.~\ref{pxpzc4}, one sees that the dashed lines have a first-order transition where $p_x/p_z$ changes discontinuously. Their manifestations are understood from the shift of the first-order transition line in Fig.~\ref{phasediagram_smallc45}.

Next, to understand the role of $c_5$, in Fig.~\ref{pxpzc5} we show the $L_xT$ dependence of $p_x/p_z$ at $T/T_{\rm d}=1.68,\ 2.10$ for three values of $c_5$ with other parameters fixed; as before, the solid lines show the results with Eq.~\eqref{FCrossParameter} and the other lines correspond to $c_5$ slightly larger or smaller than that in Eq.~\eqref{FCrossParameter}.
The phase diagram at $c_5=0.01$ is also plotted in Fig.~\ref{phasediagram_smallc45} by the dash-dotted lines. From Fig.~\ref{phasediagram_smallc45}, one finds that the point A (B) in Fig.~\ref{fig:PhaseDiagram} moves toward larger (smaller) $L_\tau T_{\rm d}$ with decreasing $c_5$. This behavior is in accordance with the fact that $\tilde f_{\rm cross}^{c_5}$ tends to trap the values of $\Omega_\tau$ and $\Omega_x$ to an intermediate value at $L_\tau=L_x$ as discussed in Sec.~\ref{sec:Ltau=Lx}. The behavior of $p_x/p_z$ in Fig.~\ref{pxpzc5} is predominantly understood from the shift of the first-order transition line,  whereas the modification of $\Omega_x$ near the first-order transition determines the precise behavior of $p_x/p_z$. 

\begin{figure}
\includegraphics[width=0.48\textwidth]{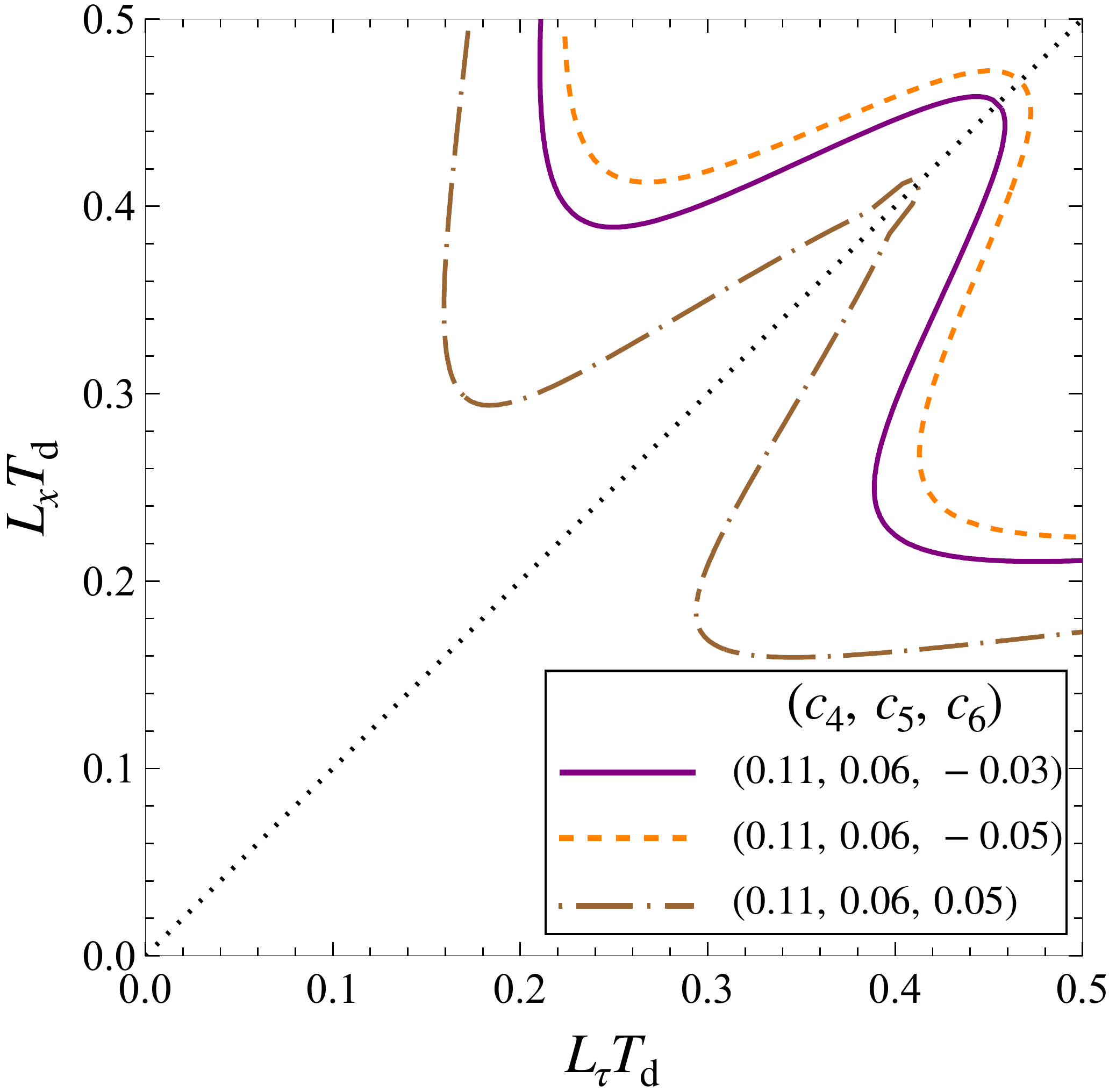}
\caption{\label{phasediagram_c6} Phase diagram on the $L_\tau$--$L_x$ plane for several values of $c_6$.}
\end{figure}

\begin{figure}
\includegraphics[scale=0.2]{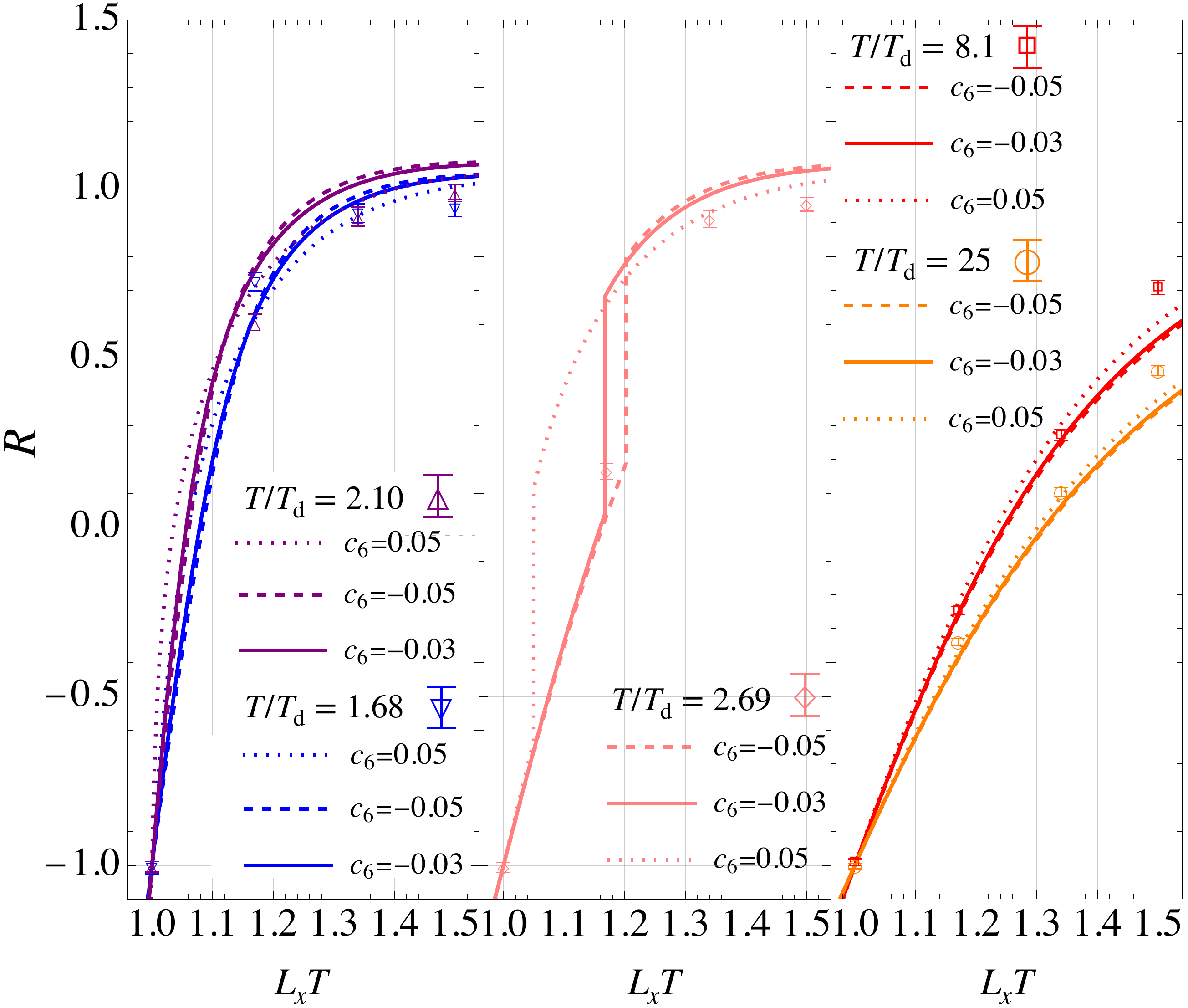}
\caption{\label{pxpzc6} $c_6$ dependence of $R$ as functions of $L_xT$ for $T/T_{\rm d}=1.68,\ 2.10,\ 2.69,\ 8.1,\ 25$. }
\end{figure}

\begin{figure}
\includegraphics[scale=0.18]{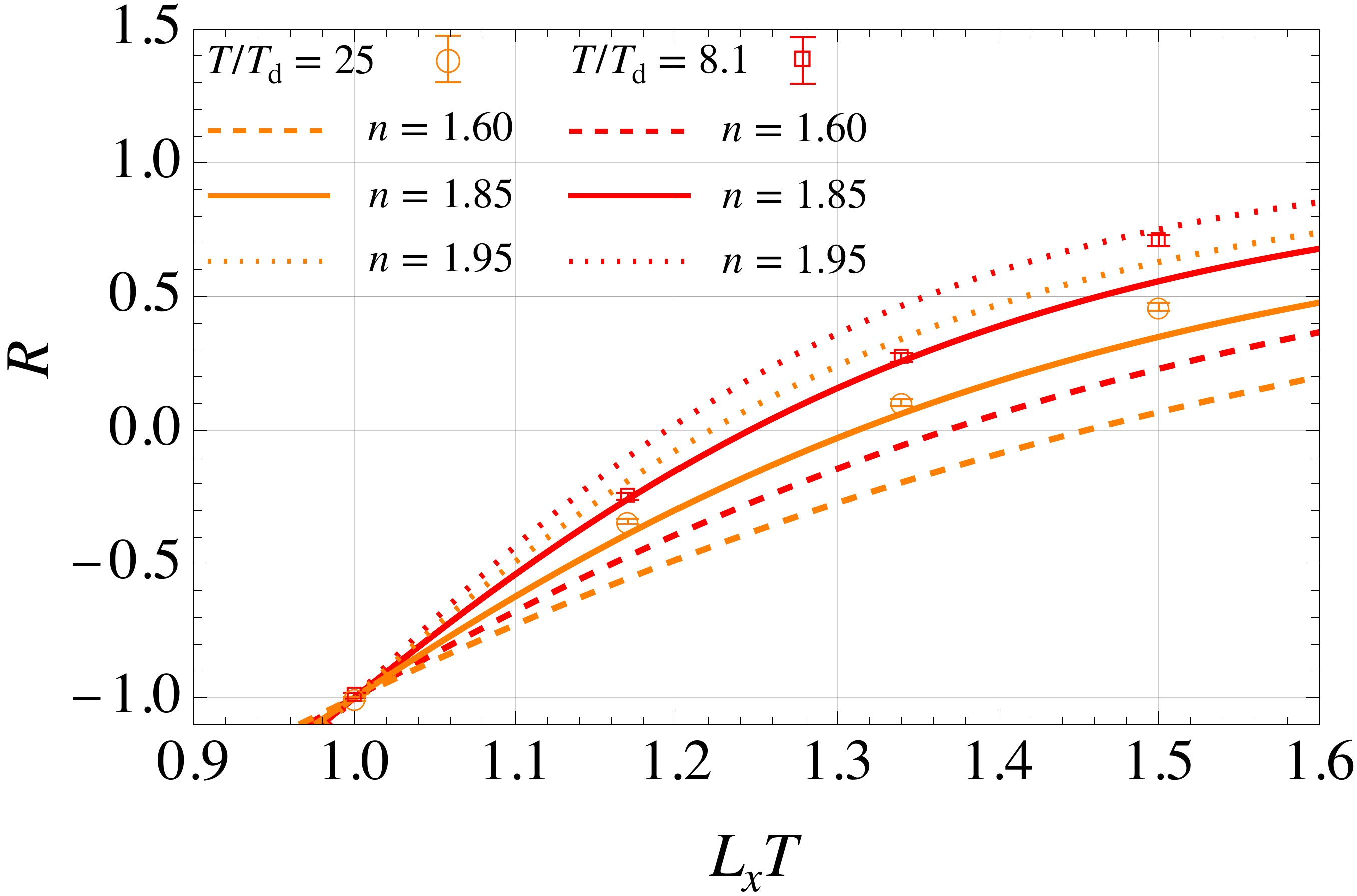}
\caption{\label{pxpzn} Dependence of $R$ on $L_xT$ for $n=1.6,\ 1.85,\ 1.95$ at $T/T_{\rm d}=8.1,25$.}
\end{figure}

In Figs.~\ref{phasediagram_c6} and~\ref{pxpzc6}, we show the phase diagram and the $L_xT$ dependence of $R$ with the variations of $c_6$ with fixed other parameters in Eq.~\eqref{FCrossParameter}. 
Figure~\ref{phasediagram_c6} shows that the location of the first-order transition in the broken phase is sensitive to this parameter. On the other hand, one finds in Fig.~\ref{pxpzc6} that the effect of this term on $R$ is small except for those caused by the shift of the first-order transition at $T/T_{\rm d}=2.69$. This result is consistent with the fact that the term in $f_{\rm cross}$ including $c_6$ is small compared with other terms as we have seen in Sec.~\ref{sec:Ltau=Lx}. This behavior indicates that $c_6$ can be used to fine-tune the location of the first-order transition without changing the overall trend of thermodynamics.

Finally, we investigate the effect of the parameter $n$. 
In Fig.~\ref{pxpzn}, we show the $L_xT$ dependence of $R$ for $T/T_{\rm d}=8.1,\ 25$ with the variation of $n$; 
the other parameters are adjusted as $(c_4,c_5,c_6)=(0.12,0.1,-0.03)$ and $(0.09,0.06,-0.03)$ for $n=1.95$ and $1.6$, respectively, to reproduce the lattice data at $T/T_{\rm d}=1.68$ and $2.10$. The solid lines show the result with Eq.~\eqref{FCrossParameter}. As the figure shows, the ratio $R$ at high temperature is sensitive to $n$, as one can easily expect from Eq.~\eqref{fTypeI}. This allows one to fix the value of $n$.

\bibliography{reference}

\begin{thebibliography}{41}%
\makeatletter
\providecommand \@ifxundefined [1]{%
 \@ifx{#1\undefined}
}%
\providecommand \@ifnum [1]{%
 \ifnum #1\expandafter \@firstoftwo
 \else \expandafter \@secondoftwo
 \fi
}%
\providecommand \@ifx [1]{%
 \ifx #1\expandafter \@firstoftwo
 \else \expandafter \@secondoftwo
 \fi
}%
\providecommand \natexlab [1]{#1}%
\providecommand \enquote  [1]{``#1''}%
\providecommand \bibnamefont  [1]{#1}%
\providecommand \bibfnamefont [1]{#1}%
\providecommand \citenamefont [1]{#1}%
\providecommand \href@noop [0]{\@secondoftwo}%
\providecommand \href [0]{\begingroup \@sanitize@url \@href}%
\providecommand \@href[1]{\@@startlink{#1}\@@href}%
\providecommand \@@href[1]{\endgroup#1\@@endlink}%
\providecommand \@sanitize@url [0]{\catcode `\\12\catcode `\$12\catcode
  `\&12\catcode `\#12\catcode `\^12\catcode `\_12\catcode `\%12\relax}%
\providecommand \@@startlink[1]{}%
\providecommand \@@endlink[0]{}%
\providecommand \url  [0]{\begingroup\@sanitize@url \@url }%
\providecommand \@url [1]{\endgroup\@href {#1}{\urlprefix }}%
\providecommand \urlprefix  [0]{URL }%
\providecommand \Eprint [0]{\href }%
\providecommand \doibase [0]{https://doi.org/}%
\providecommand \selectlanguage [0]{\@gobble}%
\providecommand \bibinfo  [0]{\@secondoftwo}%
\providecommand \bibfield  [0]{\@secondoftwo}%
\providecommand \translation [1]{[#1]}%
\providecommand \BibitemOpen [0]{}%
\providecommand \bibitemStop [0]{}%
\providecommand \bibitemNoStop [0]{.\EOS\space}%
\providecommand \EOS [0]{\spacefactor3000\relax}%
\providecommand \BibitemShut  [1]{\csname bibitem#1\endcsname}%
\let\auto@bib@innerbib\@empty
\bibitem [{\citenamefont {Boyd}\ \emph {et~al.}(1996)\citenamefont {Boyd},
  \citenamefont {Engels}, \citenamefont {Karsch}, \citenamefont {Laermann},
  \citenamefont {Legeland}, \citenamefont {Lutgemeier},\ and\ \citenamefont
  {Petersson}}]{Boyd:1996bx}%
  \BibitemOpen
  \bibfield  {author} {\bibinfo {author} {\bibfnamefont {G.}~\bibnamefont
  {Boyd}}, \bibinfo {author} {\bibfnamefont {J.}~\bibnamefont {Engels}},
  \bibinfo {author} {\bibfnamefont {F.}~\bibnamefont {Karsch}}, \bibinfo
  {author} {\bibfnamefont {E.}~\bibnamefont {Laermann}}, \bibinfo {author}
  {\bibfnamefont {C.}~\bibnamefont {Legeland}}, \bibinfo {author}
  {\bibfnamefont {M.}~\bibnamefont {Lutgemeier}},\ and\ \bibinfo {author}
  {\bibfnamefont {B.}~\bibnamefont {Petersson}},\ }\bibfield  {title} {\bibinfo
  {title} {{Thermodynamics of SU(3) lattice gauge theory}},\ }\href
  {https://doi.org/10.1016/0550-3213(96)00170-8} {\bibfield  {journal}
  {\bibinfo  {journal} {Nucl. Phys. B}\ }\textbf {\bibinfo {volume} {469}},\
  \bibinfo {pages} {419} (\bibinfo {year} {1996})},\ \Eprint
  {https://arxiv.org/abs/hep-lat/9602007} {arXiv:hep-lat/9602007} \BibitemShut
  {NoStop}%
\bibitem [{\citenamefont {Umeda}\ \emph {et~al.}(2009)\citenamefont {Umeda},
  \citenamefont {Ejiri}, \citenamefont {Aoki}, \citenamefont {Hatsuda},
  \citenamefont {Kanaya}, \citenamefont {Maezawa},\ and\ \citenamefont
  {Ohno}}]{Umeda:2008bd}%
  \BibitemOpen
  \bibfield  {author} {\bibinfo {author} {\bibfnamefont {T.}~\bibnamefont
  {Umeda}}, \bibinfo {author} {\bibfnamefont {S.}~\bibnamefont {Ejiri}},
  \bibinfo {author} {\bibfnamefont {S.}~\bibnamefont {Aoki}}, \bibinfo {author}
  {\bibfnamefont {T.}~\bibnamefont {Hatsuda}}, \bibinfo {author} {\bibfnamefont
  {K.}~\bibnamefont {Kanaya}}, \bibinfo {author} {\bibfnamefont
  {Y.}~\bibnamefont {Maezawa}},\ and\ \bibinfo {author} {\bibfnamefont
  {H.}~\bibnamefont {Ohno}},\ }\bibfield  {title} {\bibinfo {title} {{Fixed
  Scale Approach to Equation of State in Lattice QCD}},\ }\href
  {https://doi.org/10.1103/PhysRevD.79.051501} {\bibfield  {journal} {\bibinfo
  {journal} {Phys. Rev. D}\ }\textbf {\bibinfo {volume} {79}},\ \bibinfo
  {pages} {051501} (\bibinfo {year} {2009})},\ \Eprint
  {https://arxiv.org/abs/0809.2842} {arXiv:0809.2842 [hep-lat]} \BibitemShut
  {NoStop}%
\bibitem [{\citenamefont {Asakawa}\ \emph {et~al.}(2014)\citenamefont
  {Asakawa}, \citenamefont {Hatsuda}, \citenamefont {Itou}, \citenamefont
  {Kitazawa},\ and\ \citenamefont {Suzuki}}]{Asakawa:2013laa}%
  \BibitemOpen
  \bibfield  {author} {\bibinfo {author} {\bibfnamefont {M.}~\bibnamefont
  {Asakawa}}, \bibinfo {author} {\bibfnamefont {T.}~\bibnamefont {Hatsuda}},
  \bibinfo {author} {\bibfnamefont {E.}~\bibnamefont {Itou}}, \bibinfo {author}
  {\bibfnamefont {M.}~\bibnamefont {Kitazawa}},\ and\ \bibinfo {author}
  {\bibfnamefont {H.}~\bibnamefont {Suzuki}} (\bibinfo {collaboration}
  {FlowQCD}),\ }\bibfield  {title} {\bibinfo {title} {{Thermodynamics of SU(3)
  gauge theory from gradient flow on the lattice}},\ }\href
  {https://doi.org/10.1103/PhysRevD.90.011501} {\bibfield  {journal} {\bibinfo
  {journal} {Phys. Rev. D}\ }\textbf {\bibinfo {volume} {90}},\ \bibinfo
  {pages} {011501} (\bibinfo {year} {2014})},\ \bibinfo {note} {[Erratum:
  Phys.Rev.D 92, 059902 (2015)]},\ \Eprint {https://arxiv.org/abs/1312.7492}
  {arXiv:1312.7492 [hep-lat]} \BibitemShut {NoStop}%
\bibitem [{\citenamefont {Borsanyi}\ \emph {et~al.}(2012)\citenamefont
  {Borsanyi}, \citenamefont {Endrodi}, \citenamefont {Fodor}, \citenamefont
  {Katz},\ and\ \citenamefont {Szabo}}]{Borsanyi:2012ve}%
  \BibitemOpen
  \bibfield  {author} {\bibinfo {author} {\bibfnamefont {S.}~\bibnamefont
  {Borsanyi}}, \bibinfo {author} {\bibfnamefont {G.}~\bibnamefont {Endrodi}},
  \bibinfo {author} {\bibfnamefont {Z.}~\bibnamefont {Fodor}}, \bibinfo
  {author} {\bibfnamefont {S.~D.}\ \bibnamefont {Katz}},\ and\ \bibinfo
  {author} {\bibfnamefont {K.~K.}\ \bibnamefont {Szabo}},\ }\bibfield  {title}
  {\bibinfo {title} {{Precision SU(3) lattice thermodynamics for a large
  temperature range}},\ }\href {https://doi.org/10.1007/JHEP07(2012)056}
  {\bibfield  {journal} {\bibinfo  {journal} {JHEP}\ }\textbf {\bibinfo
  {volume} {07}},\ \bibinfo {pages} {056}},\ \Eprint
  {https://arxiv.org/abs/1204.6184} {arXiv:1204.6184 [hep-lat]} \BibitemShut
  {NoStop}%
\bibitem [{\citenamefont {Borsanyi}\ \emph {et~al.}(2014)\citenamefont
  {Borsanyi}, \citenamefont {Fodor}, \citenamefont {Hoelbling}, \citenamefont
  {Katz}, \citenamefont {Krieg},\ and\ \citenamefont
  {Szabo}}]{Borsanyi:2013bia}%
  \BibitemOpen
  \bibfield  {author} {\bibinfo {author} {\bibfnamefont {S.}~\bibnamefont
  {Borsanyi}}, \bibinfo {author} {\bibfnamefont {Z.}~\bibnamefont {Fodor}},
  \bibinfo {author} {\bibfnamefont {C.}~\bibnamefont {Hoelbling}}, \bibinfo
  {author} {\bibfnamefont {S.~D.}\ \bibnamefont {Katz}}, \bibinfo {author}
  {\bibfnamefont {S.}~\bibnamefont {Krieg}},\ and\ \bibinfo {author}
  {\bibfnamefont {K.~K.}\ \bibnamefont {Szabo}},\ }\bibfield  {title} {\bibinfo
  {title} {{Full result for the QCD equation of state with 2+1 flavors}},\
  }\href {https://doi.org/10.1016/j.physletb.2014.01.007} {\bibfield  {journal}
  {\bibinfo  {journal} {Phys. Lett. B}\ }\textbf {\bibinfo {volume} {730}},\
  \bibinfo {pages} {99} (\bibinfo {year} {2014})},\ \Eprint
  {https://arxiv.org/abs/1309.5258} {arXiv:1309.5258 [hep-lat]} \BibitemShut
  {NoStop}%
\bibitem [{\citenamefont {Bazavov}\ \emph {et~al.}(2014)\citenamefont {Bazavov}
  \emph {et~al.}}]{HotQCD:2014kol}%
  \BibitemOpen
  \bibfield  {author} {\bibinfo {author} {\bibfnamefont {A.}~\bibnamefont
  {Bazavov}} \emph {et~al.} (\bibinfo {collaboration} {HotQCD}),\ }\bibfield
  {title} {\bibinfo {title} {{Equation of state in ( 2+1 )-flavor QCD}},\
  }\href {https://doi.org/10.1103/PhysRevD.90.094503} {\bibfield  {journal}
  {\bibinfo  {journal} {Phys. Rev. D}\ }\textbf {\bibinfo {volume} {90}},\
  \bibinfo {pages} {094503} (\bibinfo {year} {2014})},\ \Eprint
  {https://arxiv.org/abs/1407.6387} {arXiv:1407.6387 [hep-lat]} \BibitemShut
  {NoStop}%
\bibitem [{\citenamefont {Taniguchi}\ \emph {et~al.}(2017)\citenamefont
  {Taniguchi}, \citenamefont {Ejiri}, \citenamefont {Iwami}, \citenamefont
  {Kanaya}, \citenamefont {Kitazawa}, \citenamefont {Suzuki}, \citenamefont
  {Umeda},\ and\ \citenamefont {Wakabayashi}}]{Taniguchi:2016ofw}%
  \BibitemOpen
  \bibfield  {author} {\bibinfo {author} {\bibfnamefont {Y.}~\bibnamefont
  {Taniguchi}}, \bibinfo {author} {\bibfnamefont {S.}~\bibnamefont {Ejiri}},
  \bibinfo {author} {\bibfnamefont {R.}~\bibnamefont {Iwami}}, \bibinfo
  {author} {\bibfnamefont {K.}~\bibnamefont {Kanaya}}, \bibinfo {author}
  {\bibfnamefont {M.}~\bibnamefont {Kitazawa}}, \bibinfo {author}
  {\bibfnamefont {H.}~\bibnamefont {Suzuki}}, \bibinfo {author} {\bibfnamefont
  {T.}~\bibnamefont {Umeda}},\ and\ \bibinfo {author} {\bibfnamefont
  {N.}~\bibnamefont {Wakabayashi}},\ }\bibfield  {title} {\bibinfo {title}
  {{Exploring $N_{f}$ = 2+1 QCD thermodynamics from the gradient flow}},\
  }\href {https://doi.org/10.1103/PhysRevD.96.014509} {\bibfield  {journal}
  {\bibinfo  {journal} {Phys. Rev. D}\ }\textbf {\bibinfo {volume} {96}},\
  \bibinfo {pages} {014509} (\bibinfo {year} {2017})},\ \bibinfo {note}
  {[Erratum: Phys.Rev.D 99, 059904 (2019)]},\ \Eprint
  {https://arxiv.org/abs/1609.01417} {arXiv:1609.01417 [hep-lat]} \BibitemShut
  {NoStop}%
\bibitem [{\citenamefont {Kitazawa}\ \emph {et~al.}(2016)\citenamefont
  {Kitazawa}, \citenamefont {Iritani}, \citenamefont {Asakawa}, \citenamefont
  {Hatsuda},\ and\ \citenamefont {Suzuki}}]{Kitazawa:2016dsl}%
  \BibitemOpen
  \bibfield  {author} {\bibinfo {author} {\bibfnamefont {M.}~\bibnamefont
  {Kitazawa}}, \bibinfo {author} {\bibfnamefont {T.}~\bibnamefont {Iritani}},
  \bibinfo {author} {\bibfnamefont {M.}~\bibnamefont {Asakawa}}, \bibinfo
  {author} {\bibfnamefont {T.}~\bibnamefont {Hatsuda}},\ and\ \bibinfo {author}
  {\bibfnamefont {H.}~\bibnamefont {Suzuki}},\ }\bibfield  {title} {\bibinfo
  {title} {{Equation of State for SU(3) Gauge Theory via the Energy-Momentum
  Tensor under Gradient Flow}},\ }\href
  {https://doi.org/10.1103/PhysRevD.94.114512} {\bibfield  {journal} {\bibinfo
  {journal} {Phys. Rev. D}\ }\textbf {\bibinfo {volume} {94}},\ \bibinfo
  {pages} {114512} (\bibinfo {year} {2016})},\ \Eprint
  {https://arxiv.org/abs/1610.07810} {arXiv:1610.07810 [hep-lat]} \BibitemShut
  {NoStop}%
\bibitem [{\citenamefont {Giusti}\ and\ \citenamefont
  {Pepe}(2017)}]{Giusti:2016iqr}%
  \BibitemOpen
  \bibfield  {author} {\bibinfo {author} {\bibfnamefont {L.}~\bibnamefont
  {Giusti}}\ and\ \bibinfo {author} {\bibfnamefont {M.}~\bibnamefont {Pepe}},\
  }\bibfield  {title} {\bibinfo {title} {{Equation of state of the SU(3)
  Yang\textendash{}Mills theory: A precise determination from a moving
  frame}},\ }\href {https://doi.org/10.1016/j.physletb.2017.04.001} {\bibfield
  {journal} {\bibinfo  {journal} {Phys. Lett. B}\ }\textbf {\bibinfo {volume}
  {769}},\ \bibinfo {pages} {385} (\bibinfo {year} {2017})},\ \Eprint
  {https://arxiv.org/abs/1612.00265} {arXiv:1612.00265 [hep-lat]} \BibitemShut
  {NoStop}%
\bibitem [{\citenamefont {Caselle}\ \emph {et~al.}(2018)\citenamefont
  {Caselle}, \citenamefont {Nada},\ and\ \citenamefont
  {Panero}}]{Caselle:2018kap}%
  \BibitemOpen
  \bibfield  {author} {\bibinfo {author} {\bibfnamefont {M.}~\bibnamefont
  {Caselle}}, \bibinfo {author} {\bibfnamefont {A.}~\bibnamefont {Nada}},\ and\
  \bibinfo {author} {\bibfnamefont {M.}~\bibnamefont {Panero}},\ }\bibfield
  {title} {\bibinfo {title} {{QCD thermodynamics from lattice calculations with
  nonequilibrium methods: The SU(3) equation of state}},\ }\href
  {https://doi.org/10.1103/PhysRevD.98.054513} {\bibfield  {journal} {\bibinfo
  {journal} {Phys. Rev. D}\ }\textbf {\bibinfo {volume} {98}},\ \bibinfo
  {pages} {054513} (\bibinfo {year} {2018})},\ \Eprint
  {https://arxiv.org/abs/1801.03110} {arXiv:1801.03110 [hep-lat]} \BibitemShut
  {NoStop}%
\bibitem [{\citenamefont {Iritani}\ \emph {et~al.}(2019)\citenamefont
  {Iritani}, \citenamefont {Kitazawa}, \citenamefont {Suzuki},\ and\
  \citenamefont {Takaura}}]{Iritani:2018idk}%
  \BibitemOpen
  \bibfield  {author} {\bibinfo {author} {\bibfnamefont {T.}~\bibnamefont
  {Iritani}}, \bibinfo {author} {\bibfnamefont {M.}~\bibnamefont {Kitazawa}},
  \bibinfo {author} {\bibfnamefont {H.}~\bibnamefont {Suzuki}},\ and\ \bibinfo
  {author} {\bibfnamefont {H.}~\bibnamefont {Takaura}},\ }\bibfield  {title}
  {\bibinfo {title} {{Thermodynamics in quenched QCD: energy--momentum tensor
  with two-loop order coefficients in the gradient flow formalism}},\ }\href
  {https://doi.org/10.1093/ptep/ptz001} {\bibfield  {journal} {\bibinfo
  {journal} {PTEP}\ }\textbf {\bibinfo {volume} {2019}},\ \bibinfo {pages}
  {023B02} (\bibinfo {year} {2019})},\ \Eprint
  {https://arxiv.org/abs/1812.06444} {arXiv:1812.06444 [hep-lat]} \BibitemShut
  {NoStop}%
\bibitem [{\citenamefont {Yagi}\ \emph {et~al.}(2005)\citenamefont {Yagi},
  \citenamefont {Hatsuda},\ and\ \citenamefont {Miake}}]{Yagi:2005yb}%
  \BibitemOpen
  \bibfield  {author} {\bibinfo {author} {\bibfnamefont {K.}~\bibnamefont
  {Yagi}}, \bibinfo {author} {\bibfnamefont {T.}~\bibnamefont {Hatsuda}},\ and\
  \bibinfo {author} {\bibfnamefont {Y.}~\bibnamefont {Miake}},\ }\href@noop {}
  {\emph {\bibinfo {title} {{Quark-gluon plasma: From big bang to little
  bang}}}},\ Vol.~\bibinfo {volume} {23}\ (\bibinfo {year} {2005})\BibitemShut
  {NoStop}%
\bibitem [{\citenamefont {Casimir}(1948)}]{Casimir:1948dh}%
  \BibitemOpen
  \bibfield  {author} {\bibinfo {author} {\bibfnamefont {H.~B.~G.}\
  \bibnamefont {Casimir}},\ }\bibfield  {title} {\bibinfo {title} {{On the
  Attraction Between Two Perfectly Conducting Plates}},\ }\href@noop {}
  {\bibfield  {journal} {\bibinfo  {journal} {Indag. Math.}\ }\textbf {\bibinfo
  {volume} {10}},\ \bibinfo {pages} {261} (\bibinfo {year} {1948})}\BibitemShut
  {NoStop}%
\bibitem [{\citenamefont {Brown}\ and\ \citenamefont
  {Maclay}(1969)}]{Brown:1969na}%
  \BibitemOpen
  \bibfield  {author} {\bibinfo {author} {\bibfnamefont {L.~S.}\ \bibnamefont
  {Brown}}\ and\ \bibinfo {author} {\bibfnamefont {G.~J.}\ \bibnamefont
  {Maclay}},\ }\bibfield  {title} {\bibinfo {title} {{Vacuum stress between
  conducting plates: An Image solution}},\ }\href
  {https://doi.org/10.1103/PhysRev.184.1272} {\bibfield  {journal} {\bibinfo
  {journal} {Phys. Rev.}\ }\textbf {\bibinfo {volume} {184}},\ \bibinfo {pages}
  {1272} (\bibinfo {year} {1969})}\BibitemShut {NoStop}%
\bibitem [{\citenamefont {Hanada}\ and\ \citenamefont
  {Kanamori}(2009)}]{Hanada:2009hq}%
  \BibitemOpen
  \bibfield  {author} {\bibinfo {author} {\bibfnamefont {M.}~\bibnamefont
  {Hanada}}\ and\ \bibinfo {author} {\bibfnamefont {I.}~\bibnamefont
  {Kanamori}},\ }\bibfield  {title} {\bibinfo {title} {{Lattice study of
  two-dimensional N=(2,2) super Yang-Mills at large-N}},\ }\href
  {https://doi.org/10.1103/PhysRevD.80.065014} {\bibfield  {journal} {\bibinfo
  {journal} {Phys. Rev. D}\ }\textbf {\bibinfo {volume} {80}},\ \bibinfo
  {pages} {065014} (\bibinfo {year} {2009})},\ \Eprint
  {https://arxiv.org/abs/0907.4966} {arXiv:0907.4966 [hep-lat]} \BibitemShut
  {NoStop}%
\bibitem [{\citenamefont {Hanada}\ \emph {et~al.}(2011)\citenamefont {Hanada},
  \citenamefont {Matsuura},\ and\ \citenamefont {Sugino}}]{Hanada:2010kt}%
  \BibitemOpen
  \bibfield  {author} {\bibinfo {author} {\bibfnamefont {M.}~\bibnamefont
  {Hanada}}, \bibinfo {author} {\bibfnamefont {S.}~\bibnamefont {Matsuura}},\
  and\ \bibinfo {author} {\bibfnamefont {F.}~\bibnamefont {Sugino}},\
  }\bibfield  {title} {\bibinfo {title} {{Two-dimensional lattice for
  four-dimensional N=4 supersymmetric Yang-Mills}},\ }\href
  {https://doi.org/10.1143/PTP.126.597} {\bibfield  {journal} {\bibinfo
  {journal} {Prog. Theor. Phys.}\ }\textbf {\bibinfo {volume} {126}},\ \bibinfo
  {pages} {597} (\bibinfo {year} {2011})},\ \Eprint
  {https://arxiv.org/abs/1004.5513} {arXiv:1004.5513 [hep-lat]} \BibitemShut
  {NoStop}%
\bibitem [{\citenamefont {\"Unsal}\ and\ \citenamefont
  {Yaffe}(2010)}]{Unsal:2010qh}%
  \BibitemOpen
  \bibfield  {author} {\bibinfo {author} {\bibfnamefont {M.}~\bibnamefont
  {\"Unsal}}\ and\ \bibinfo {author} {\bibfnamefont {L.~G.}\ \bibnamefont
  {Yaffe}},\ }\bibfield  {title} {\bibinfo {title} {{Large-N volume
  independence in conformal and confining gauge theories}},\ }\href
  {https://doi.org/10.1007/JHEP08(2010)030} {\bibfield  {journal} {\bibinfo
  {journal} {JHEP}\ }\textbf {\bibinfo {volume} {08}},\ \bibinfo {pages}
  {030}},\ \Eprint {https://arxiv.org/abs/1006.2101} {arXiv:1006.2101 [hep-th]}
  \BibitemShut {NoStop}%
\bibitem [{\citenamefont {Mandal}\ and\ \citenamefont
  {Morita}(2011)}]{Mandal:2011hb}%
  \BibitemOpen
  \bibfield  {author} {\bibinfo {author} {\bibfnamefont {G.}~\bibnamefont
  {Mandal}}\ and\ \bibinfo {author} {\bibfnamefont {T.}~\bibnamefont
  {Morita}},\ }\bibfield  {title} {\bibinfo {title} {{Phases of a two
  dimensional large N gauge theory on a torus}},\ }\href
  {https://doi.org/10.1103/PhysRevD.84.085007} {\bibfield  {journal} {\bibinfo
  {journal} {Phys. Rev. D}\ }\textbf {\bibinfo {volume} {84}},\ \bibinfo
  {pages} {085007} (\bibinfo {year} {2011})},\ \Eprint
  {https://arxiv.org/abs/1103.1558} {arXiv:1103.1558 [hep-th]} \BibitemShut
  {NoStop}%
\bibitem [{\citenamefont {Mandal}\ and\ \citenamefont
  {Morita}(2013)}]{Mandal:2013id}%
  \BibitemOpen
  \bibfield  {author} {\bibinfo {author} {\bibfnamefont {G.}~\bibnamefont
  {Mandal}}\ and\ \bibinfo {author} {\bibfnamefont {T.}~\bibnamefont
  {Morita}},\ }\bibfield  {title} {\bibinfo {title} {{Quantum quench in matrix
  models: Dynamical phase transitions, Selective equilibration and the
  Generalized Gibbs Ensemble}},\ }\href
  {https://doi.org/10.1007/JHEP10(2013)197} {\bibfield  {journal} {\bibinfo
  {journal} {JHEP}\ }\textbf {\bibinfo {volume} {10}},\ \bibinfo {pages}
  {197}},\ \Eprint {https://arxiv.org/abs/1302.0859} {arXiv:1302.0859 [hep-th]}
  \BibitemShut {NoStop}%
\bibitem [{\citenamefont {Ishikawa}\ \emph
  {et~al.}(2019{\natexlab{a}})\citenamefont {Ishikawa}, \citenamefont
  {Nakayama},\ and\ \citenamefont {Suzuki}}]{Ishikawa:2018yey}%
  \BibitemOpen
  \bibfield  {author} {\bibinfo {author} {\bibfnamefont {T.}~\bibnamefont
  {Ishikawa}}, \bibinfo {author} {\bibfnamefont {K.}~\bibnamefont {Nakayama}},\
  and\ \bibinfo {author} {\bibfnamefont {K.}~\bibnamefont {Suzuki}},\
  }\bibfield  {title} {\bibinfo {title} {{Casimir effect for nucleon parity
  doublets}},\ }\href {https://doi.org/10.1103/PhysRevD.99.054010} {\bibfield
  {journal} {\bibinfo  {journal} {Phys. Rev. D}\ }\textbf {\bibinfo {volume}
  {99}},\ \bibinfo {pages} {054010} (\bibinfo {year} {2019}{\natexlab{a}})},\
  \Eprint {https://arxiv.org/abs/1812.10964} {arXiv:1812.10964 [hep-ph]}
  \BibitemShut {NoStop}%
\bibitem [{\citenamefont {Ishikawa}\ \emph
  {et~al.}(2019{\natexlab{b}})\citenamefont {Ishikawa}, \citenamefont
  {Nakayama}, \citenamefont {Suenaga},\ and\ \citenamefont
  {Suzuki}}]{Ishikawa:2019dcn}%
  \BibitemOpen
  \bibfield  {author} {\bibinfo {author} {\bibfnamefont {T.}~\bibnamefont
  {Ishikawa}}, \bibinfo {author} {\bibfnamefont {K.}~\bibnamefont {Nakayama}},
  \bibinfo {author} {\bibfnamefont {D.}~\bibnamefont {Suenaga}},\ and\ \bibinfo
  {author} {\bibfnamefont {K.}~\bibnamefont {Suzuki}},\ }\bibfield  {title}
  {\bibinfo {title} {{$D$ mesons as a probe of Casimir effect for chiral
  symmetry breaking}},\ }\href {https://doi.org/10.1103/PhysRevD.100.034016}
  {\bibfield  {journal} {\bibinfo  {journal} {Phys. Rev. D}\ }\textbf {\bibinfo
  {volume} {100}},\ \bibinfo {pages} {034016} (\bibinfo {year}
  {2019}{\natexlab{b}})},\ \Eprint {https://arxiv.org/abs/1905.11164}
  {arXiv:1905.11164 [hep-ph]} \BibitemShut {NoStop}%
\bibitem [{\citenamefont {Kitazawa}\ \emph {et~al.}(2019)\citenamefont
  {Kitazawa}, \citenamefont {Mogliacci}, \citenamefont {Kolb\'e},\ and\
  \citenamefont {Horowitz}}]{Kitazawa:2019otp}%
  \BibitemOpen
  \bibfield  {author} {\bibinfo {author} {\bibfnamefont {M.}~\bibnamefont
  {Kitazawa}}, \bibinfo {author} {\bibfnamefont {S.}~\bibnamefont {Mogliacci}},
  \bibinfo {author} {\bibfnamefont {I.}~\bibnamefont {Kolb\'e}},\ and\ \bibinfo
  {author} {\bibfnamefont {W.~A.}\ \bibnamefont {Horowitz}},\ }\bibfield
  {title} {\bibinfo {title} {{Anisotropic pressure induced by finite-size
  effects in SU(3) Yang-Mills theory}},\ }\href
  {https://doi.org/10.1103/PhysRevD.99.094507} {\bibfield  {journal} {\bibinfo
  {journal} {Phys. Rev. D}\ }\textbf {\bibinfo {volume} {99}},\ \bibinfo
  {pages} {094507} (\bibinfo {year} {2019})},\ \Eprint
  {https://arxiv.org/abs/1904.00241} {arXiv:1904.00241 [hep-lat]} \BibitemShut
  {NoStop}%
\bibitem [{\citenamefont {Inagaki}\ \emph {et~al.}(2022)\citenamefont
  {Inagaki}, \citenamefont {Matsuo},\ and\ \citenamefont
  {Shimoji}}]{Inagaki:2021yhi}%
  \BibitemOpen
  \bibfield  {author} {\bibinfo {author} {\bibfnamefont {T.}~\bibnamefont
  {Inagaki}}, \bibinfo {author} {\bibfnamefont {Y.}~\bibnamefont {Matsuo}},\
  and\ \bibinfo {author} {\bibfnamefont {H.}~\bibnamefont {Shimoji}},\
  }\bibfield  {title} {\bibinfo {title} {{Precise phase structure in a
  four-fermion interaction model on a torus}},\ }\href
  {https://doi.org/10.1093/ptep/ptab160} {\bibfield  {journal} {\bibinfo
  {journal} {PTEP}\ }\textbf {\bibinfo {volume} {2022}},\ \bibinfo {pages}
  {013B09} (\bibinfo {year} {2022})},\ \Eprint
  {https://arxiv.org/abs/2108.03583} {arXiv:2108.03583 [hep-ph]} \BibitemShut
  {NoStop}%
\bibitem [{\citenamefont {Chernodub}\ \emph {et~al.}(2022)\citenamefont
  {Chernodub}, \citenamefont {Goy}, \citenamefont {Molochkov},\ and\
  \citenamefont {Tanashkin}}]{Chernodub:2022izt}%
  \BibitemOpen
  \bibfield  {author} {\bibinfo {author} {\bibfnamefont {M.~N.}\ \bibnamefont
  {Chernodub}}, \bibinfo {author} {\bibfnamefont {V.~A.}\ \bibnamefont {Goy}},
  \bibinfo {author} {\bibfnamefont {A.~V.}\ \bibnamefont {Molochkov}},\ and\
  \bibinfo {author} {\bibfnamefont {A.~S.}\ \bibnamefont {Tanashkin}},\
  }\bibfield  {title} {\bibinfo {title} {{Casimir boundaries, monopoles, and
  deconfinement transition in (3+1)- dimensional compact electrodynamics}},\
  }\href {https://doi.org/10.1103/PhysRevD.105.114506} {\bibfield  {journal}
  {\bibinfo  {journal} {Phys. Rev. D}\ }\textbf {\bibinfo {volume} {105}},\
  \bibinfo {pages} {114506} (\bibinfo {year} {2022})},\ \Eprint
  {https://arxiv.org/abs/2203.14922} {arXiv:2203.14922 [hep-lat]} \BibitemShut
  {NoStop}%
\bibitem [{\citenamefont {Tanizaki}\ and\ \citenamefont
  {\"Unsal}(2022{\natexlab{a}})}]{Tanizaki:2022plm}%
  \BibitemOpen
  \bibfield  {author} {\bibinfo {author} {\bibfnamefont {Y.}~\bibnamefont
  {Tanizaki}}\ and\ \bibinfo {author} {\bibfnamefont {M.}~\bibnamefont
  {\"Unsal}},\ }\bibfield  {title} {\bibinfo {title} {{Semiclassics with
  \textquoteright{}t Hooft flux background for QCD with 2-index quarks}},\
  }\href {https://doi.org/10.1007/JHEP08(2022)038} {\bibfield  {journal}
  {\bibinfo  {journal} {JHEP}\ }\textbf {\bibinfo {volume} {08}},\ \bibinfo
  {pages} {038}},\ \Eprint {https://arxiv.org/abs/2205.11339} {arXiv:2205.11339
  [hep-th]} \BibitemShut {NoStop}%
\bibitem [{\citenamefont {Tanizaki}\ and\ \citenamefont
  {\"Unsal}(2022{\natexlab{b}})}]{Tanizaki:2022ngt}%
  \BibitemOpen
  \bibfield  {author} {\bibinfo {author} {\bibfnamefont {Y.}~\bibnamefont
  {Tanizaki}}\ and\ \bibinfo {author} {\bibfnamefont {M.}~\bibnamefont
  {\"Unsal}},\ }\bibfield  {title} {\bibinfo {title} {{Center vortex and
  confinement in Yang\textendash{}Mills theory and QCD with anomaly-preserving
  compactifications}},\ }\href {https://doi.org/10.1093/ptep/ptac042}
  {\bibfield  {journal} {\bibinfo  {journal} {PTEP}\ }\textbf {\bibinfo
  {volume} {2022}},\ \bibinfo {pages} {04A108} (\bibinfo {year}
  {2022}{\natexlab{b}})},\ \Eprint {https://arxiv.org/abs/2201.06166}
  {arXiv:2201.06166 [hep-th]} \BibitemShut {NoStop}%
\bibitem [{\citenamefont {Hayashi}\ \emph {et~al.}(2023)\citenamefont
  {Hayashi}, \citenamefont {Tanizaki},\ and\ \citenamefont
  {Watanabe}}]{Hayashi:2023wwi}%
  \BibitemOpen
  \bibfield  {author} {\bibinfo {author} {\bibfnamefont {Y.}~\bibnamefont
  {Hayashi}}, \bibinfo {author} {\bibfnamefont {Y.}~\bibnamefont {Tanizaki}},\
  and\ \bibinfo {author} {\bibfnamefont {H.}~\bibnamefont {Watanabe}},\
  }\bibfield  {title} {\bibinfo {title} {{Semiclassical analysis of the
  bifundamental QCD on~$\mathbb{R}^2\times T^2$ with \textquoteright{}t Hooft
  flux}},\ }\href {https://doi.org/10.1007/JHEP10(2023)146} {\bibfield
  {journal} {\bibinfo  {journal} {JHEP}\ }\textbf {\bibinfo {volume} {10}},\
  \bibinfo {pages} {146}},\ \Eprint {https://arxiv.org/abs/2307.13954}
  {arXiv:2307.13954 [hep-th]} \BibitemShut {NoStop}%
\bibitem [{\citenamefont {Hayashi}\ and\ \citenamefont
  {Tanizaki}(2024)}]{Hayashi:2024qkm}%
  \BibitemOpen
  \bibfield  {author} {\bibinfo {author} {\bibfnamefont {Y.}~\bibnamefont
  {Hayashi}}\ and\ \bibinfo {author} {\bibfnamefont {Y.}~\bibnamefont
  {Tanizaki}},\ }\bibfield  {title} {\bibinfo {title} {{Semiclassics for the
  QCD vacuum structure through T$^{2}$-compactification with the
  baryon-\textquoteright{}t Hooft flux}},\ }\href
  {https://doi.org/10.1007/JHEP08(2024)001} {\bibfield  {journal} {\bibinfo
  {journal} {JHEP}\ }\textbf {\bibinfo {volume} {08}},\ \bibinfo {pages}
  {001}},\ \Eprint {https://arxiv.org/abs/2402.04320} {arXiv:2402.04320
  [hep-th]} \BibitemShut {NoStop}%
\bibitem [{\citenamefont {Meisinger}\ \emph {et~al.}(2002)\citenamefont
  {Meisinger}, \citenamefont {Miller},\ and\ \citenamefont
  {Ogilvie}}]{Meisinger:2001cq}%
  \BibitemOpen
  \bibfield  {author} {\bibinfo {author} {\bibfnamefont {P.~N.}\ \bibnamefont
  {Meisinger}}, \bibinfo {author} {\bibfnamefont {T.~R.}\ \bibnamefont
  {Miller}},\ and\ \bibinfo {author} {\bibfnamefont {M.~C.}\ \bibnamefont
  {Ogilvie}},\ }\bibfield  {title} {\bibinfo {title} {{Phenomenological
  equations of state for the quark gluon plasma}},\ }\href
  {https://doi.org/10.1103/PhysRevD.65.034009} {\bibfield  {journal} {\bibinfo
  {journal} {Phys. Rev. D}\ }\textbf {\bibinfo {volume} {65}},\ \bibinfo
  {pages} {034009} (\bibinfo {year} {2002})},\ \Eprint
  {https://arxiv.org/abs/hep-ph/0108009} {arXiv:hep-ph/0108009} \BibitemShut
  {NoStop}%
\bibitem [{\citenamefont {Pisarski}(2000)}]{Pisarski:2000eq}%
  \BibitemOpen
  \bibfield  {author} {\bibinfo {author} {\bibfnamefont {R.~D.}\ \bibnamefont
  {Pisarski}},\ }\bibfield  {title} {\bibinfo {title} {{Quark gluon plasma as a
  condensate of Z(3) Wilson lines}},\ }\href
  {https://doi.org/10.1103/PhysRevD.62.111501} {\bibfield  {journal} {\bibinfo
  {journal} {Phys. Rev. D}\ }\textbf {\bibinfo {volume} {62}},\ \bibinfo
  {pages} {111501} (\bibinfo {year} {2000})},\ \Eprint
  {https://arxiv.org/abs/hep-ph/0006205} {arXiv:hep-ph/0006205} \BibitemShut
  {NoStop}%
\bibitem [{\citenamefont {Dumitru}\ \emph {et~al.}(2011)\citenamefont
  {Dumitru}, \citenamefont {Guo}, \citenamefont {Hidaka}, \citenamefont
  {Altes},\ and\ \citenamefont {Pisarski}}]{Dumitru:2010mj}%
  \BibitemOpen
  \bibfield  {author} {\bibinfo {author} {\bibfnamefont {A.}~\bibnamefont
  {Dumitru}}, \bibinfo {author} {\bibfnamefont {Y.}~\bibnamefont {Guo}},
  \bibinfo {author} {\bibfnamefont {Y.}~\bibnamefont {Hidaka}}, \bibinfo
  {author} {\bibfnamefont {C.~P.~K.}\ \bibnamefont {Altes}},\ and\ \bibinfo
  {author} {\bibfnamefont {R.~D.}\ \bibnamefont {Pisarski}},\ }\bibfield
  {title} {\bibinfo {title} {{How Wide is the Transition to Deconfinement?}},\
  }\href {https://doi.org/10.1103/PhysRevD.83.034022} {\bibfield  {journal}
  {\bibinfo  {journal} {Phys. Rev. D}\ }\textbf {\bibinfo {volume} {83}},\
  \bibinfo {pages} {034022} (\bibinfo {year} {2011})},\ \Eprint
  {https://arxiv.org/abs/1011.3820} {arXiv:1011.3820 [hep-ph]} \BibitemShut
  {NoStop}%
\bibitem [{\citenamefont {Dumitru}\ \emph {et~al.}(2012)\citenamefont
  {Dumitru}, \citenamefont {Guo}, \citenamefont {Hidaka}, \citenamefont
  {Altes},\ and\ \citenamefont {Pisarski}}]{Dumitru:2012fw}%
  \BibitemOpen
  \bibfield  {author} {\bibinfo {author} {\bibfnamefont {A.}~\bibnamefont
  {Dumitru}}, \bibinfo {author} {\bibfnamefont {Y.}~\bibnamefont {Guo}},
  \bibinfo {author} {\bibfnamefont {Y.}~\bibnamefont {Hidaka}}, \bibinfo
  {author} {\bibfnamefont {C.~P.~K.}\ \bibnamefont {Altes}},\ and\ \bibinfo
  {author} {\bibfnamefont {R.~D.}\ \bibnamefont {Pisarski}},\ }\bibfield
  {title} {\bibinfo {title} {{Effective Matrix Model for Deconfinement in Pure
  Gauge Theories}},\ }\href {https://doi.org/10.1103/PhysRevD.86.105017}
  {\bibfield  {journal} {\bibinfo  {journal} {Phys. Rev. D}\ }\textbf {\bibinfo
  {volume} {86}},\ \bibinfo {pages} {105017} (\bibinfo {year} {2012})},\
  \Eprint {https://arxiv.org/abs/1205.0137} {arXiv:1205.0137 [hep-ph]}
  \BibitemShut {NoStop}%
\bibitem [{\citenamefont {Fukushima}\ and\ \citenamefont
  {Skokov}(2017)}]{Fukushima:2017csk}%
  \BibitemOpen
  \bibfield  {author} {\bibinfo {author} {\bibfnamefont {K.}~\bibnamefont
  {Fukushima}}\ and\ \bibinfo {author} {\bibfnamefont {V.}~\bibnamefont
  {Skokov}},\ }\bibfield  {title} {\bibinfo {title} {{Polyakov loop modeling
  for hot QCD}},\ }\href {https://doi.org/10.1016/j.ppnp.2017.05.002}
  {\bibfield  {journal} {\bibinfo  {journal} {Prog. Part. Nucl. Phys.}\
  }\textbf {\bibinfo {volume} {96}},\ \bibinfo {pages} {154} (\bibinfo {year}
  {2017})},\ \Eprint {https://arxiv.org/abs/1705.00718} {arXiv:1705.00718
  [hep-ph]} \BibitemShut {NoStop}%
\bibitem [{\citenamefont {Suenaga}\ and\ \citenamefont
  {Kitazawa}(2023)}]{Suenaga:2022rjk}%
  \BibitemOpen
  \bibfield  {author} {\bibinfo {author} {\bibfnamefont {D.}~\bibnamefont
  {Suenaga}}\ and\ \bibinfo {author} {\bibfnamefont {M.}~\bibnamefont
  {Kitazawa}},\ }\bibfield  {title} {\bibinfo {title} {{Effective model for
  pure Yang-Mills theory on
  ${\mathbb{T}}^{2}\ifmmode\times\else\texttimes\fi{}{\mathbb{R}}^{2}$ with
  Polyakov loops}},\ }\href {https://doi.org/10.1103/PhysRevD.107.074502}
  {\bibfield  {journal} {\bibinfo  {journal} {Phys. Rev. D}\ }\textbf {\bibinfo
  {volume} {107}},\ \bibinfo {pages} {074502} (\bibinfo {year} {2023})},\
  \Eprint {https://arxiv.org/abs/2210.09363} {arXiv:2210.09363 [hep-ph]}
  \BibitemShut {NoStop}%
\bibitem [{\citenamefont {Rothe}(2012)}]{Rothe:1992nt}%
  \BibitemOpen
  \bibfield  {author} {\bibinfo {author} {\bibfnamefont {H.~J.}\ \bibnamefont
  {Rothe}},\ }\href {https://doi.org/10.1142/8229} {\emph {\bibinfo {title}
  {{Lattice Gauge Theories : An Introduction (Fourth Edition)}}}},\
  Vol.~\bibinfo {volume} {43}\ (\bibinfo  {publisher} {World Scientific
  Publishing Company},\ \bibinfo {year} {2012})\BibitemShut {NoStop}%
\bibitem [{\citenamefont {Sasaki}\ \emph {et~al.}(2014)\citenamefont {Sasaki},
  \citenamefont {Mishustin},\ and\ \citenamefont {Redlich}}]{Sasaki:2013xfa}%
  \BibitemOpen
  \bibfield  {author} {\bibinfo {author} {\bibfnamefont {C.}~\bibnamefont
  {Sasaki}}, \bibinfo {author} {\bibfnamefont {I.}~\bibnamefont {Mishustin}},\
  and\ \bibinfo {author} {\bibfnamefont {K.}~\bibnamefont {Redlich}},\
  }\bibfield  {title} {\bibinfo {title} {{Implementation of chromomagnetic
  gluons in Yang-Mills thermodynamics}},\ }\href
  {https://doi.org/10.1103/PhysRevD.89.014031} {\bibfield  {journal} {\bibinfo
  {journal} {Phys. Rev. D}\ }\textbf {\bibinfo {volume} {89}},\ \bibinfo
  {pages} {014031} (\bibinfo {year} {2014})},\ \Eprint
  {https://arxiv.org/abs/1308.3635} {arXiv:1308.3635 [hep-ph]} \BibitemShut
  {NoStop}%
\bibitem [{\citenamefont {Asakawa}\ and\ \citenamefont
  {Yazaki}(1989)}]{Asakawa:1989bq}%
  \BibitemOpen
  \bibfield  {author} {\bibinfo {author} {\bibfnamefont {M.}~\bibnamefont
  {Asakawa}}\ and\ \bibinfo {author} {\bibfnamefont {K.}~\bibnamefont
  {Yazaki}},\ }\bibfield  {title} {\bibinfo {title} {{Chiral Restoration at
  Finite Density and Temperature}},\ }\href
  {https://doi.org/10.1016/0375-9474(89)90002-X} {\bibfield  {journal}
  {\bibinfo  {journal} {Nucl. Phys. A}\ }\textbf {\bibinfo {volume} {504}},\
  \bibinfo {pages} {668} (\bibinfo {year} {1989})}\BibitemShut {NoStop}%
\bibitem [{\citenamefont {Kitazawa}\ \emph {et~al.}(2002)\citenamefont
  {Kitazawa}, \citenamefont {Koide}, \citenamefont {Kunihiro},\ and\
  \citenamefont {Nemoto}}]{Kitazawa:2002jop}%
  \BibitemOpen
  \bibfield  {author} {\bibinfo {author} {\bibfnamefont {M.}~\bibnamefont
  {Kitazawa}}, \bibinfo {author} {\bibfnamefont {T.}~\bibnamefont {Koide}},
  \bibinfo {author} {\bibfnamefont {T.}~\bibnamefont {Kunihiro}},\ and\
  \bibinfo {author} {\bibfnamefont {Y.}~\bibnamefont {Nemoto}},\ }\bibfield
  {title} {\bibinfo {title} {{Chiral and color superconducting phase
  transitions with vector interaction in a simple model}},\ }\href
  {https://doi.org/10.1143/PTP.108.929} {\bibfield  {journal} {\bibinfo
  {journal} {Prog. Theor. Phys.}\ }\textbf {\bibinfo {volume} {108}},\ \bibinfo
  {pages} {929} (\bibinfo {year} {2002})},\ \bibinfo {note} {[Erratum:
  Prog.Theor.Phys. 110, 185--186 (2003)]},\ \Eprint
  {https://arxiv.org/abs/hep-ph/0207255} {arXiv:hep-ph/0207255} \BibitemShut
  {NoStop}%
\bibitem [{\citenamefont {Philipsen}(2021)}]{Philipsen:2021qji}%
  \BibitemOpen
  \bibfield  {author} {\bibinfo {author} {\bibfnamefont {O.}~\bibnamefont
  {Philipsen}},\ }\bibfield  {title} {\bibinfo {title} {{Lattice Constraints on
  the QCD Chiral Phase Transition at Finite Temperature and Baryon Density}},\
  }\href {https://doi.org/10.3390/sym13112079} {\bibfield  {journal} {\bibinfo
  {journal} {Symmetry}\ }\textbf {\bibinfo {volume} {13}},\ \bibinfo {pages}
  {2079} (\bibinfo {year} {2021})},\ \Eprint {https://arxiv.org/abs/2111.03590}
  {arXiv:2111.03590 [hep-lat]} \BibitemShut {NoStop}%
\bibitem [{\citenamefont {Chernodub}\ \emph {et~al.}(2017)\citenamefont
  {Chernodub}, \citenamefont {Goy},\ and\ \citenamefont
  {Molochkov}}]{Chernodub:2017mhi}%
  \BibitemOpen
  \bibfield  {author} {\bibinfo {author} {\bibfnamefont {M.~N.}\ \bibnamefont
  {Chernodub}}, \bibinfo {author} {\bibfnamefont {V.~A.}\ \bibnamefont {Goy}},\
  and\ \bibinfo {author} {\bibfnamefont {A.~V.}\ \bibnamefont {Molochkov}},\
  }\bibfield  {title} {\bibinfo {title} {{Nonperturbative Casimir effect and
  monopoles: compact Abelian gauge theory in two spatial dimensions}},\ }\href
  {https://doi.org/10.1103/PhysRevD.95.074511} {\bibfield  {journal} {\bibinfo
  {journal} {Phys. Rev. D}\ }\textbf {\bibinfo {volume} {95}},\ \bibinfo
  {pages} {074511} (\bibinfo {year} {2017})},\ \Eprint
  {https://arxiv.org/abs/1703.03439} {arXiv:1703.03439 [hep-lat]} \BibitemShut
  {NoStop}%
\bibitem [{\citenamefont {Chernodub}\ \emph {et~al.}(2019)\citenamefont
  {Chernodub}, \citenamefont {Goy},\ and\ \citenamefont
  {Molochkov}}]{Chernodub:2018aix}%
  \BibitemOpen
  \bibfield  {author} {\bibinfo {author} {\bibfnamefont {M.~N.}\ \bibnamefont
  {Chernodub}}, \bibinfo {author} {\bibfnamefont {V.~A.}\ \bibnamefont {Goy}},\
  and\ \bibinfo {author} {\bibfnamefont {A.~V.}\ \bibnamefont {Molochkov}},\
  }\bibfield  {title} {\bibinfo {title} {{Phase structure of lattice Yang-Mills
  theory on ${\mathbb T}^2 \times {\mathbb R}^2$}},\ }\href
  {https://doi.org/10.1103/PhysRevD.99.074021} {\bibfield  {journal} {\bibinfo
  {journal} {Phys. Rev. D}\ }\textbf {\bibinfo {volume} {99}},\ \bibinfo
  {pages} {074021} (\bibinfo {year} {2019})},\ \Eprint
  {https://arxiv.org/abs/1811.01550} {arXiv:1811.01550 [hep-lat]} \BibitemShut
  {NoStop}%
\end{thebibliography}%

\end{document}